\documentclass[a4paper,11pt]{article}
\usepackage{graphicx}
\usepackage{epsf,amsmath,bbold,amsfonts,stmaryrd}

\usepackage[utf8]{inputenc}
\usepackage{mathrsfs}
\usepackage{appendix}
\usepackage{amssymb}
\usepackage{float}
\usepackage{color}
\usepackage{cite}
\usepackage{hyperref}
\hypersetup{pageanchor=false}
\usepackage{indentfirst}
\usepackage{url}
\usepackage{float}
\usepackage{caption}
\usepackage[numbers,square,comma,sort&compress,merge]{natbib}

\usepackage{subfigure}

\def\nn{\nonumber}

\def\be{\begin{equation}}
\def\ee{\end{equation}}

\def\bea{\begin{eqnarray}}
\def\eea{\end{eqnarray}}

\def\ba{\begin{array}}
\def\ea{\end{array}}

\def\bc{\begin{center}}
\def\ec{\end{center}}

\def\bl{\begin{flushleft}}
\def\el{\end{flushleft}}

\def\br{\begin{flushright}}
\def\er{\end{flushright}}

\def\bi{\begin{itemize}}
\def\ei{\end{itemize}}

\def\bt{\begin{tabular}}
\def\et{\end{tabular}}

\numberwithin{equation}{section}

\begin{document}
\title{\textbf{QED Effects on Kerr Black Hole Shadows immersed in uniform magnetic fields}}
\author{Zhen Zhong$^{1}$, Zezhou Hu$^{2}$, Haopeng Yan$^3$, Minyong Guo$^{1*}$ and Bin Chen$^{2,3,4}$}
\date{}
\maketitle

\vspace{-10mm}

\begin{center}
{\it
$^1$Department of Physics, Beijing Normal University, Beijing 100875, P. R. China\\\vspace{1mm}

$^2$School of Physics, Peking University, No.5 Yiheyuan Rd, Beijing
100871, P.R. China\\\vspace{1mm}

$^3$Center for High Energy Physics, Peking University,
No.5 Yiheyuan Rd, Beijing 100871, P. R. China\\\vspace{1mm}

$^4$Collaborative Innovation Center of Quantum Matter, No.5 Yiheyuan Rd, Beijing 100871, P. R.
China

}
\end{center}

\vspace{8mm}

\begin{abstract}
In this work, taking the QED effect into account, we investigate the shadows of the Kerr black holes immersed in uniform magnetic fields through the numerical backward ray-tracing method. We introduce a dimensionless parameter $\Lambda$ to characterize the strength of magnetic fields and study the influence of magnetic fields on the Kerr black hole shadows for various spins of the black holes and inclination angles of the observers. In particular, we find that the photon ``hairs''  appear near the left edge of the shadow in the presence of magnetic fields. The photon hairs may be served as a signature of the magnetic fields. We notice that the photon  hairs become more evident when the strength of magnetic fields or the spin of the black hole becomes larger. In addition, we study the deformation of the shadows by bringing in quantitative parameters that can describe the position and shape of the shadow edge.

\end{abstract}

\vfill{\footnotesize $^*$Corresponding author: minyongguo@bnu.edu.cn.}

\maketitle

\newpage

\section{Introduction}
Black holes play a very important role in understanding the nature of gravity and spacetimes. As technology improves, people are now able to hear black holes through LIGO and Virgo \cite{LIGOScientific:2016aoc}, and to see black holes with the help of the Event Horizon Telescope (EHT) \cite{EventHorizonTelescope:2019dse}.  In particular, the image of the supermassive black hole at the center of M87 galaxy has the up-down asymmetry, that is, a prominent rotationally symmetric mode, which suggests that the black hole has to be spinning. Up to now, the best model mimicking a real rotating black hole in our universe is generally considered to be the Kerr spacetime. With the confidence and motivation given by the observational images of black holes, theoretical research has been widely carried out in recently years. Inspired by the pioneer works \cite{Synge:1966okc, Bardeen:1973tla}, people have studied the observational signatures of supermassive objects on many aspects including the shadows \cite{Cunha:2015yba, Wei:2019pjf, Li:2020drn, Cunha:2016bjh, Guo:2020zmf, Perlick:2021aok, Wang:2016wcj, Wang:2018eui, Hou:2021okc, KIczek:2021vlc, Cardoso:2019rvt, Contreras:2021yxe, Chowdhuri:2020ipb}, the images of companion stars \cite{Gralla:2017ufe, Guo:2018kis, Yan:2019etp, Guo:2019lur}, the photon rings \cite{Gralla:2019xty, Peng:2021osd, Johannsen:2013vgc, Guo:2020qwk, Zeng:2020dco, Zeng:2021dlj, Gan:2021pwu, Peng:2020wun} and etc. \cite{Zhang:2019glo, Cai:2021fpr, Han:2018ooi, Wang:2020emr, Guo:2019pte, Li:2020val, Yan:2021yuo, Cardoso:2008bp, Konoplya:2007yy, Yang:2021zqy, Li:2021zct, Konoplya:2006qr, Konoplya:2006gg}.

Recently, almost two years after the first picture was released in 2019, EHT Collaboration published the polarized image of M$87^*$, from which we can clearly see that there is an extra twisting polarization pattern for the bright ring compared to the first image \cite{EventHorizonTelescope:2021bee}. Since in magnetic fields the polarization of photons would rotate as in Faraday effects, it has been argued that the polarized image should be caused by the synchrotron radiation of innermost charged particles of the accretion disk moving in a magnetic field. Thus the the polarized image of M$87^*$ indicates that, around the black hole, there must exist a strong magnetic field \cite{EventHorizonTelescope:2021srq} and the polarization structure depends on both the spacetime and the magnetic fields around a black hole. 

As we know, classically the magnetic fields do not affect the trajectories of photons, so that black hole shadows formed by countless light traces are only directly determined by the spacetime structure. However, in practice there are a few indirect ways through which the magnetic fields can  affect the motion of the photons. One obvious way is that if the magnetic field is strong enough, its reaction to the spacetime should be taken seriously such that the background metric would be changed. Consequently the motions of the photons get modified as well.  Along this line, in \cite{Lima:2021cgb} and \cite{Wang:2021ara}, the authors have investigated the Schwarzschild  and the Kerr black hole shadows in the Melvin magnetic field. The advantage of this analysis is that the influence of the magnetic field on the black hole shadows can be easily studied. Whereas, the spacetime in a Melvin magnetic field is not asymptotically flat any more. 

In this paper, we would like to study the influence of the magnetic fields on the Kerr black hole shadow in an alternative way, which has been discussed in our previous paper \cite{Hu:2020usx} in the context of a static black hole. As we know, astronomical black holes in the reality are  always expected to be rotating and the Kerr black hole solution has been considered as the most competitive model to mimic a real black hole in the universe. Thus, it is necessary to generalize our previous work \cite{Hu:2020usx} to the Kerr black holes. In addition, compared to the  Schwarzschild black holes, the shadows of the Kerr black holes have much richer structures. Therefore, it is also interesting to see how magnetic fields affect the shadows when the spin of a black hole becomes significant. Moreover, as mentioned above, we pay our attention to the situation that the magnetic fields are not strong enough to modify the geometry of background spacetime, instead, our strategy is to take quantum electrodynamics (QED) effect on the photons into account. In this case, by including the birefringence \cite{DeLorenci:2000yh, Novello:1999pg} induced by the QED effect, the motions of massless particles would deviate from the geodesics if the black hole is immersed in a magnetic field, and now we are able to investigate the effects of magnetic fields on the Kerr black hole shadows.  In this work, we pay our attention to the Kerr black hole with a uniform magnetic field outside the horizon. Using the numerical backward ray-tracing method, we investigate the Kerr black hole shadows under the influence of the magnetic field. In particular, we show that with increasing magnetic fields or increasing spin, the photon hairs appearing near the left  edge of the shadow, which could be taken as a signature of the magnetic field, become more evident. Furthermore, we introduce six parameters to characterize the edges of the shadows, and quantitatively calculate the deformations of the shadow edges affected by the uniform magnetic field.

The remaining parts of this paper is organized as follows. In section \ref{section2}, we give a brief review on the dispersion relations induced by the QED effect and derive the equations of motion. In section \ref{section3}, we study the shadows of Kerr black holes immersed in the uniform magnetic fields in detail. In section \ref{summary}, we summarize our results. In this work, we have set the fundamental constants $c$, $G$, the vacuum permittivity $\varepsilon_0$ and the mass of the black hole $M$ to unity, and we will work in the convention $(-, +, +, +)$.

\section{Dispersion relations and equations of motion}\label{section2}
As shown in our previous paper \cite{Hu:2020usx}, the action involving the electromagnetic gauge potential minimally coupled to gravity takes the form
\bea
I=\int d^4\sqrt{-g}\left(\frac{1}{16\pi}R+L_{\text{eff}}\right),
\eea
where $L_{\text{eff}}$ is the Euler-Heisenberg effective Lagrangian for the electromagnetic field which can describe the one-loop vacuum polarization and reads
\bea
  L_{\text{eff}} = - \frac{1}{4} F_{\mu \nu} F^{\mu \nu} - \frac{\mu}{2} \left[ \frac{5}{4} (F_{\mu \nu} F^{\mu \nu})^2 - \frac{7}{2} F_{\mu
  \nu} F_{\sigma \tau} F^{\mu \sigma} F^{\nu \tau} \right],
\eea
and the coupling constant is defined as
\bea\label{mu}
\mu=\frac{\hbar e^4}{360\pi^2m_e^4},
\eea
where $m_e$ is the electron mass  and $e$ denotes the charge of the electron. After straightforward and standard calculations, one can obtain the dispersion relation and the effective metric given as follows,
\bea\label{effectivemetric}
0&=&p_\alpha p_\beta\left(g^{\alpha\beta}+X_{\alpha\beta}\right),\label{null}\\
G_{\alpha \beta}&=& g_{\alpha \beta} + X_{\alpha\beta},\label{effg}
\eea
where $p_\alpha$ is the momentum of the photon, and  a new tensor $X_{\alpha\beta}\equiv \lambda F^{\mu}_{~~\alpha}F_{\mu\beta}$ was introduced for simplicity with $\lambda = - 8 \mu$ or $- 14 \mu$ giving two different polarizations of the photons.

With the above effective metric, one can define the dual vector $q_\mu$ as $G_{\mu\nu} p^\nu$. Then the conjugate variables $(x^\mu , q_\mu)$ would satisfy the canonical equations as follows,
\be\label{eee}
\dot{q}_\mu=-\frac{\partial H}{\partial x^\mu},  \quad \dot{x}^\mu=\frac{\partial H}{\partial q_\mu},
\ee
\be\label{eee2}
H=H(q_\mu,x^\mu)=G^{\mu\nu}(x) q_\mu q_\nu.
\ee
We want to stress that here we use $G^{\mu \nu}$ to denote the inverse matrix of $G_{\mu \nu}$, other than $G^{\mu\nu}(x)= g^{\mu\rho} g^{\nu\sigma} G_{\rho \sigma}$.  That is, we have $G^{\mu\nu} G_{\nu \sigma}\equiv\delta^{\mu}_{\sigma}$.

\section{Shadow of Kerr black holes in uniform magnetic field}\label{section3}

In this section, we focus on a specific case that the Kerr black hole is immersed in the uniform magnetic field. The uniform magnetic field we are interested in is assumed to be not very strong, so that we are allowed to ignore the back-reaction from the magnetic field to the Kerr metric. The Kerr metric in the Boyer-Lindquist coordinate takes the form
\begin{equation}
    g_{\mu \nu}d x^{\mu}d x^{\nu}=-(1-\frac{2 r}{\Sigma})d t^2+\frac{\Sigma}{\Delta} d r^2+\Sigma d \theta^2+\frac{1}{\Sigma}[(r^2+a^2)^2-\Delta a^2 \sin^2 \theta] \sin^2 \theta d \phi^2-\frac{4 a r}{\Sigma}\sin^2\theta d t d\phi.
\end{equation}
where
\be
\Delta=r^2-2 r+a^2,\qquad
\Sigma=r^2+a^2\cos^2\theta.
\ee
We consider a stationary uniform magnetic field, which is independent of the coordinates $t$ and $\phi$. The gauge field $A_\mu$ can be solved from the source-free Maxwell equations, that is, $\nabla_\mu F^{\mu\nu}=0$. An elegant procedure to find the solutions was first presented in \cite{Wald:1974np}. And, the non-zero components of the gauge field $A_\mu$ read
\bea
    A_t&=&-\frac{a B r}{\Sigma} \sin^2 \theta-2 a B (\frac{1}{2}-\frac{r}{\Sigma}),\nn\\
    A_\phi&=&\frac{B}{\Sigma}[(r^2+a^2)^2-\Delta a^2 \sin^2 \theta] \sin^2 \theta - \frac{2 a^2 B r}{\Sigma} \sin^2 \theta.
\eea
Here in our convation we have set the mass of the black hole to be unity, hence $a$ can be directly treated as the angular momentum of the Kerr black hole.

Then from Eq. (\ref{effg}), we can obtain the effective metric. However, the explicit forms of the components of $G_{\mu\nu}$ are too complicated to be written down here, and we leave them into the appendix(\ref{appendixA}). Based on the effective metric, we would like to determine the motions of the photons using the Hamilton-Jacobi equation
\be
G^{\mu\nu}(x) \frac{\partial S}{\partial x^\mu} \frac{\partial S}{\partial x^\nu}=0.
\ee
After some calculations, one can easily see that the system is non-integrable due to the lack of the symmetry associated with the Carter constant. Thus in order to figure out the shadow structure of the effective black hole, we have to turn to the numerical backward ray-tracing method which has been introduced in detail in the paper \cite{Hu:2020usx}.  
In the present work, we would not like to bore the readers with the details of the numerical backward ray-tracing method. Instead, let us review the essential points of the method. 
\begin{figure}[h!]

  \centering
  \includegraphics[width=3.5in]{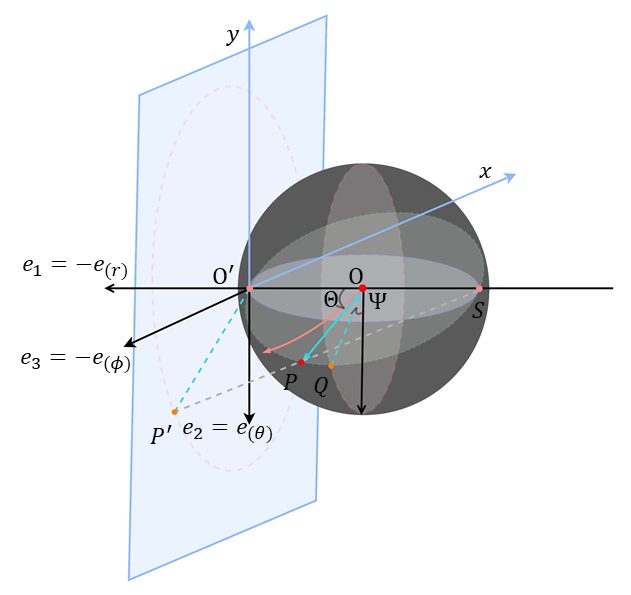}

  \caption{The ZAMO tetrad of the camera and the celestial coordinates $\Theta$ and $\Phi$ based on the stereographic projection. This diagram has been shown in Fig. 11 of our previous work \cite{Hu:2020usx}.}
  \label{zamo}
\end{figure}

Roughly speaking, we use spherical illumination to obtain the image of the black hole in the backward ray-tracing method. We place a ball-like extended source in the infinity to illuminate the system. And the black hole and the camera are inside the ball-like source.  Furthermore, we divide the ball into the grids with the network made of latitudinal and longitudinal lines. Each grid is assigned with a color standing for four symmetric parts of the celestial sphere. In addition, we employ the fish-eye camera model to photograph the Kerr black hole. The local rest frame of the camera we use is the usual zero-angular-momentum-observer (ZAMO) tetrad which takes the following form:
\bea
e_0&=&\frac{g_{\phi\phi}\partial_t-g_{\phi t}\partial_{\phi}}{\sqrt{g_{\phi\phi}\left(g_{\phi t}^2-g_{\phi\phi}g_{tt}\right)}},\\
e_1&=&-\frac{\partial_r}{\sqrt{g_{rr}}},\\
e_2&=&\frac{\partial_\theta}{\sqrt{g_{\theta\theta}}},\\
e_3&=&-\frac{\partial_\phi}{\sqrt{g_{\phi\phi}}}.
\eea
In the ZAMO frame, the electromagnetic field in the background can be written as
\bea
X_{(a)(b)}=X_{\mu\nu}e^\mu_{(a)}e^\nu_{(b)}\label{zamoe}.
\eea
On the other hand, we use the stereographic projection to build a connection between the celestial sphere to the screen of our camera, and the celestial coordinates $\Theta$ and $\Phi$ can be understood from Fig. \ref{zamo}. Thus, in the ZAMO frame, the $4$-momentum of a photon can be expressed as
\bea
p=-\kappa e_{(0)}+|OP|\left(\cos\Theta e_{(1)}+\sin\Theta \cos\Psi e_{(2)}+\sin\Theta \sin\Psi e_{(3)}\right).\label{zamop}
\eea
Note that the photons travel along the null geodesics with respect to the effective metric. That is, from the Eq. (\ref{null}), one has
\bea
-\kappa^2+|OP|^2+\kappa^2 c_1+\kappa |OP|c_2+|OP|^2 c_3=0\label{kappa},
\eea
where combining with Eqs. (\ref{zamoe}) and (\ref{zamop}), the factors $c_i$ ($i=1,2,3$) read
\bea
c_1&=&\gamma\left(X_{tt}+\frac{g_{t\phi}^2}{g_{\phi\phi}^2}X_{\phi\phi}
-2\frac{g_{t\phi}}{g_{\phi\phi}}X_{t\phi}\right),\\
c_2&=&2\sqrt{\gamma}\left(\frac{\cos\Theta}{\sqrt{g_{rr}}}\zeta_r
-\frac{\sin\Theta\cos\Psi}{\sqrt{g_{\theta\theta}}}\zeta_\theta
+\frac{\sin\Theta\sin\Psi}{\sqrt{g_{\phi\phi}}}\zeta_\phi\right),\\
c_3&=&\frac{X_{rr}}{g_{rr}}\cos^2\Theta-\frac{2X_{r\theta}}{\sqrt{g_{rr}g_{\theta\theta}}}\cos\Theta\sin\Theta\cos\Psi+\frac{X_{\theta\theta}}{g_{\theta\theta}}\sin^2\Theta\cos^2\Psi\nn\\
&&+\frac{2X_{r\phi}}{\sqrt{g_{rr}g_{\phi\phi}}}\cos\Theta\sin
\Theta\sin\Psi-\frac{2X_{\theta\phi}}{\sqrt{g_{\theta\theta}
g_{\phi\phi}}}\sin^2\Theta\sin\Psi\cos\Psi+\frac{X_{\phi\phi}}
{g_{\phi\phi}}\sin^2\Theta\sin^2\Psi,
\eea
in which we have also introduced several new parameters, including
\bea
\gamma=\frac{g_{\phi\phi}}{g_{t\phi}-g_{tt}g_{\phi\phi}}>0,
\eea
outside the horizon of the Kerr black hole, and
\bea
\zeta_i=X_{ti}-\frac{g_{t\phi}}{g_{\phi\phi}}X_{i\phi}, \quad i=r, \theta, \phi,
\eea
appeared in the expression of $c_2$. Then by solving the quadratic equation (\ref{kappa}), the factor $\kappa$ can be determined to be 
\bea
\kappa=\frac{c_2+\sqrt{4+c_2^2+4c_3-4c_1(1+c_3)}}{2(1-c_1)}|OP|. 
\eea
Here we have excluded the negative root of Eq. (\ref{kappa}), since we need $\kappa>0$ to ensure that the $4$-momentum of the photon is a past-directed vector to employ the backward ray-tracing method. Moreover, using the stereographic projection, we can build a map from the celestial sphere to the screen of our camera. In addition, note that,  the momentum vector can be also expressed in the coordinate bases, that is, $p^\mu=\frac{dx^\mu}{d\tau}$. Thus, compared with Eq. (\ref{zamop}), one can have the values of $q_\mu$ for the given pixels on the screen of the camera. Then, combining with the coordinates of the camera, one can find sets of $(x^\mu, q_\mu)$. Putting these sets as initial values, we can do the numerical geodesic evolutions using Eqs. (\ref{eee}) and (\ref{eee2}), and then identify the photons that will not fall into the black hole.

\subsection{Deformation of Kerr black hole shadow}
From the appendix \ref{appendixA}, we can see that in the effective metric, among the three parameters $a$, $\lambda$ and $B$, the latter two always show up together. More precisely, $\lambda$ is always partnered with $B^2$ in the effective metric and the geodesic equations. Hence,  we can define a dimensionless parameter $\Lambda$ standing for the field strength as
\be
\Lambda=\lambda B^2.
\ee

In addition, when performing the calculations of black hole shadow, we should guarantee the validness of the causality of the effective metric. In the case at hand, we should always make sure $G_{\theta\theta}, G_{\phi\phi}>0$ outside the horizon and $G_{tt}<0$ outside the ergoregion. Unfortunately, from the expressions of the components of the effective metric, we can not find an analytical bound for $\Lambda$. Therefore, we numerically check these inequalities and find that at least in the range $-0.2\le\Lambda\le0$, the effective metric works well. Thus, in the following, we would like to pay our attention to the allowed range $-0.2\le\Lambda\le0$. 

Before we get into the analysis of shadows, we would like to take a little space to discuss how large the magnetic field is in International System of Units (SI for short) if we say $-0.2\le\Lambda\le0$ in our convention $c=G=\epsilon_0=M=1$. From Eq. (\ref{mu}), we can find
\bea
\mu=\frac{3.7\times10^{68}}{M_{BH}^2}\,,
\eea 
after some dimensional analysis, where $\mu$ is the value of the coupling constant in our convension and $M_{BH}$ is the value of the black hole mass in SI. For example, $\mu=0.1$ corresponds to $M_{BH}\simeq6\times10^{34}\,\text{kg}=3\times10^4 M_{\odot}$ with $M_{\odot}$ being the mass of sun. On the other hand, for the magnetic fields, Gauss unit rather than SI is customarily used. Similarly, we can find
\bea
B=6\times10^{-51} B_{GS}M_{BH}\,,
\eea
where $B$ is the the value of  magnetic field intensity in our convension, $M_{BH}$ is the value in SI, and $B_{GS}$ is the value of the magnetic field intensity in Gauss unit. Then, we have 
\bea
\Lambda=\lambda B^2\simeq-B_{GS}^2\times10^{30}\,.
\eea
From this relation, we can see that $-0.2\le\Lambda\le0$ means the magnetic field intensity is approximately $0\leq B_{GS}\leq  4.5\times10^{14}\text{Gauss}$.

Now we are ready to explore the shadows of the Kerr black holes bathed in the uniform magnetic fields in various cases. Firstly, let us focus on the the variations of the shadows with respect to the spin $a$ of the Kerr black hole with fixed $\Lambda$ and $\theta_o$, where $\theta_o$ is the inclination angle of the observer.

\begin{figure}[h!]
  \centering

  \subfigure[$a=0.2$]{
  \begin{minipage}[t]{0.3\linewidth}
  \centering
  \includegraphics[width=1.5in]{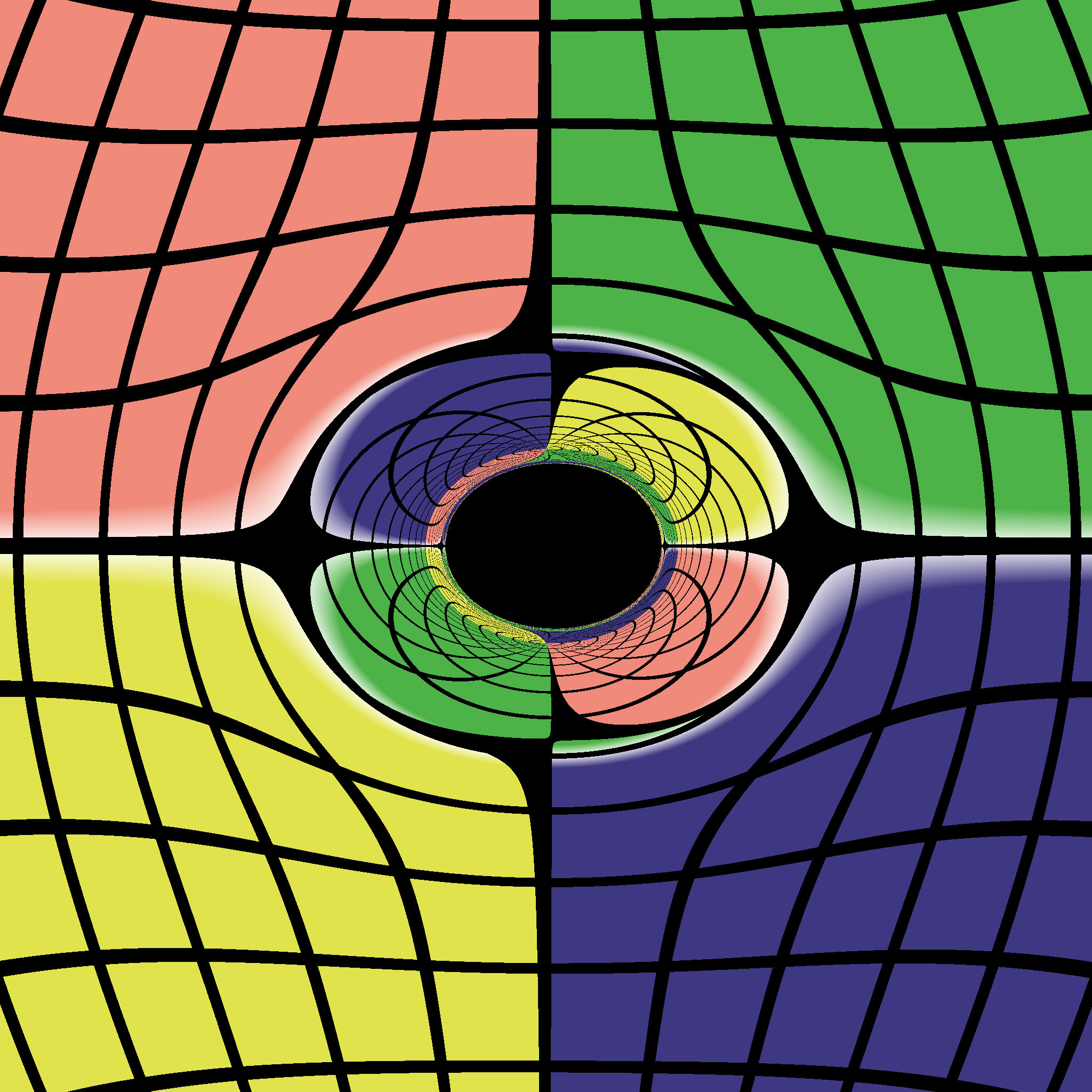}
  \end{minipage}
  }
  \subfigure[$a=0.5$]{
  \begin{minipage}[t]{0.3\linewidth}
  \centering
  \includegraphics[width=1.5in]{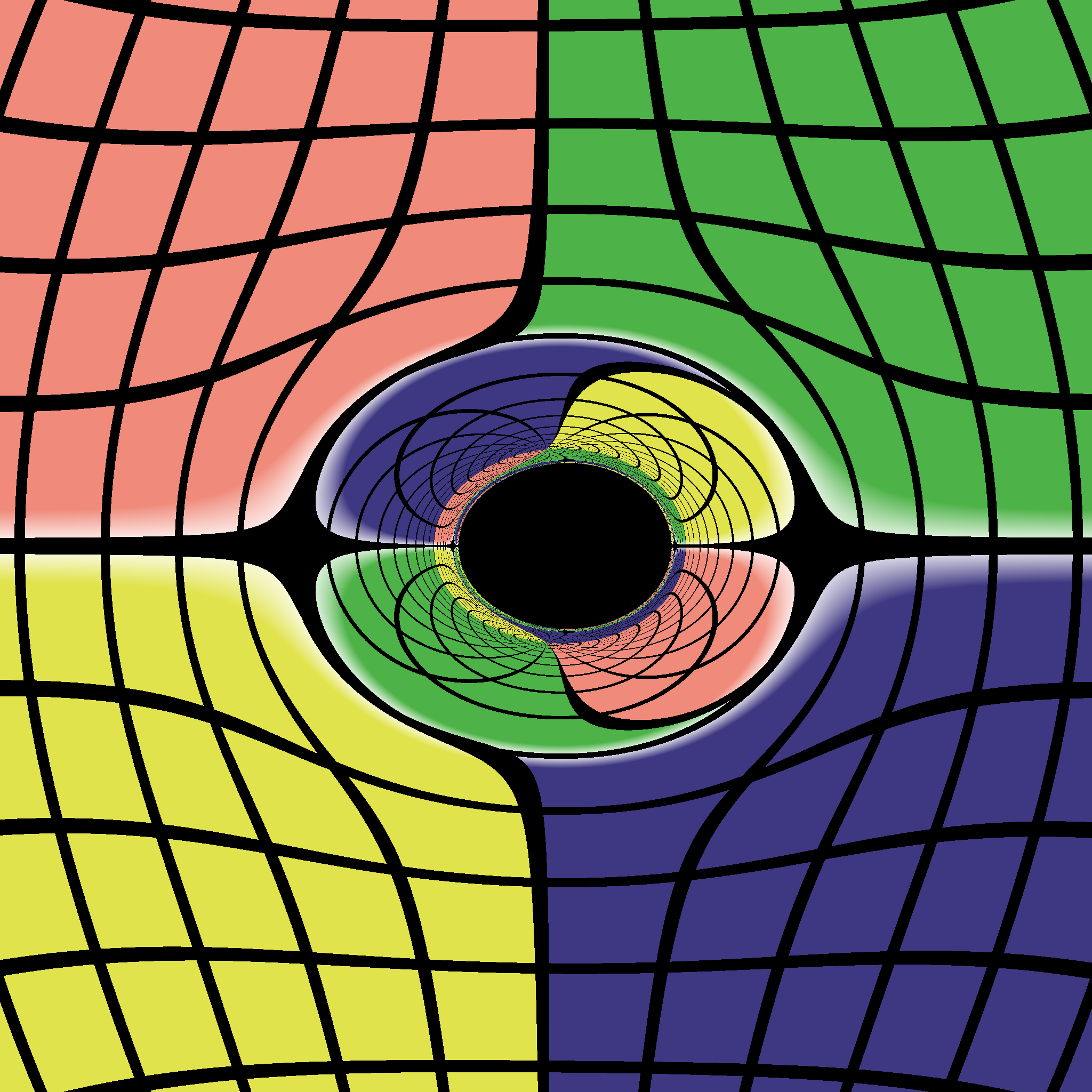}
  \end{minipage}%
  }%
\subfigure[$a=0.7$]{
  \begin{minipage}[t]{0.3\linewidth}
  \centering
  \includegraphics[width=1.5in]{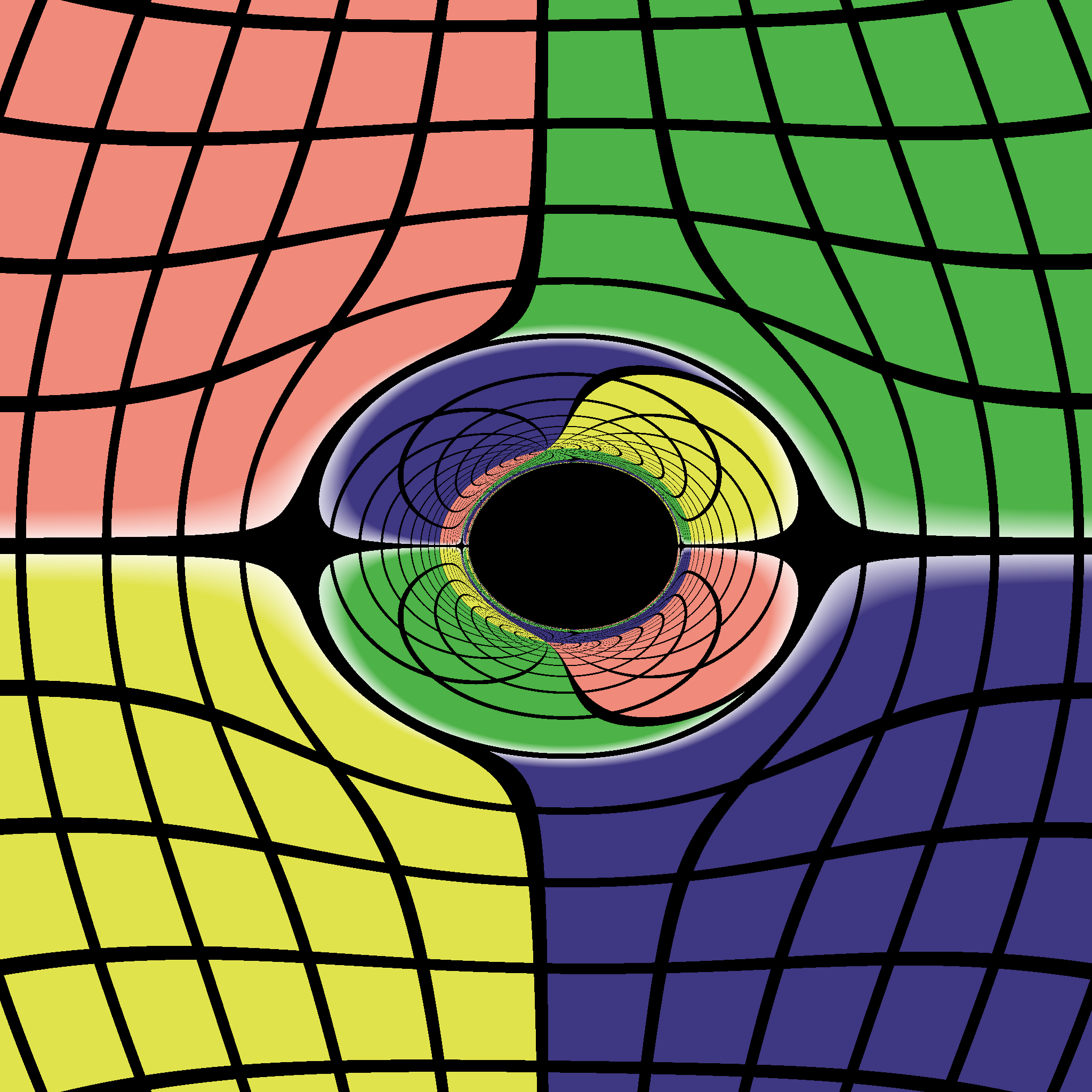}
  \end{minipage}
  }%

  \subfigure[$a=0.9$]{
  \begin{minipage}[t]{0.3\linewidth}
  \centering
  \includegraphics[width=1.5in]{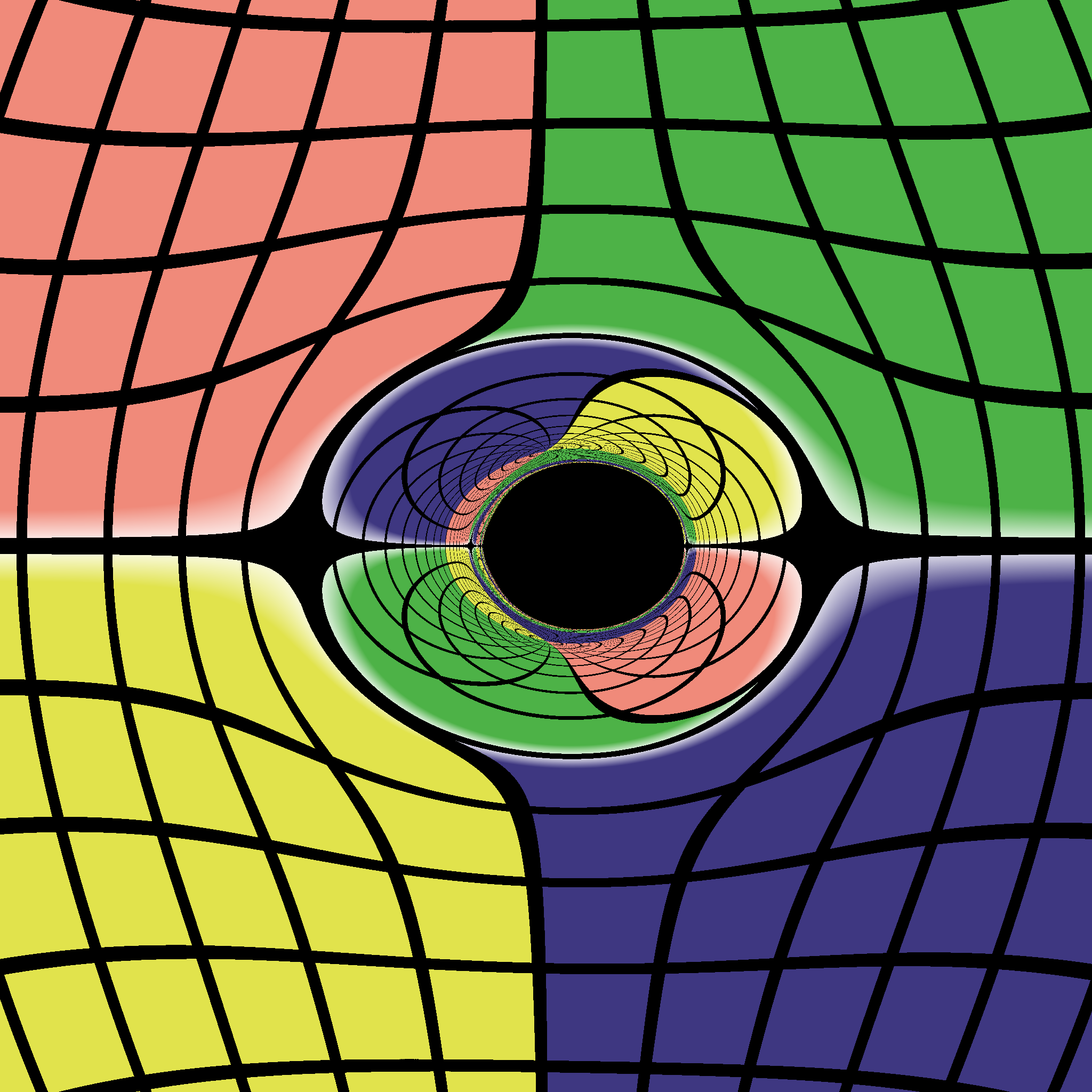}
  \end{minipage}
  }%
  \subfigure[$a=0.99$]{
  \begin{minipage}[t]{0.3\linewidth}
  \centering
  \includegraphics[width=1.5in]{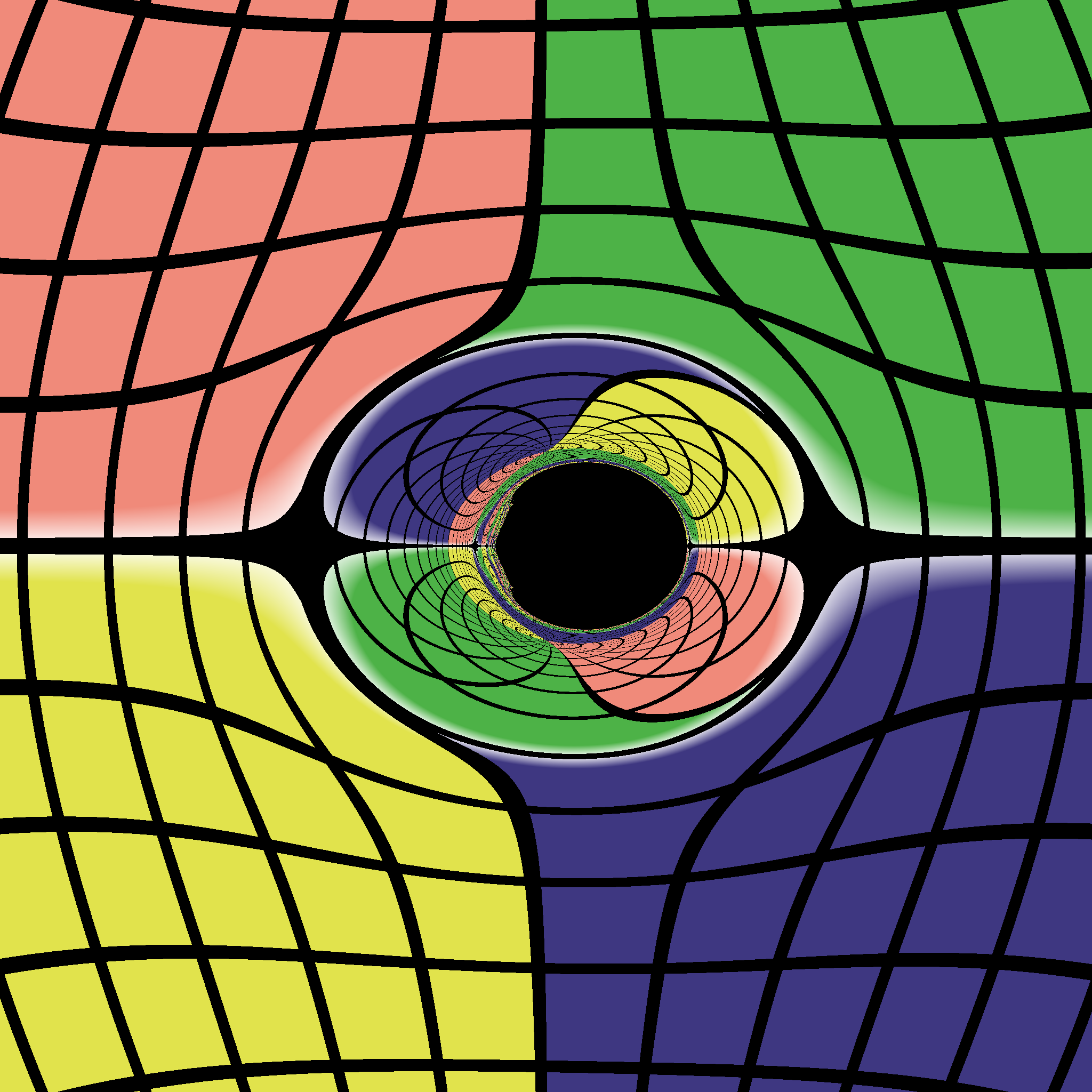}
  \end{minipage}
  }%
  \subfigure[$a=0.999$]{
  \begin{minipage}[t]{0.3\linewidth}
  \centering
  \includegraphics[width=1.5in]{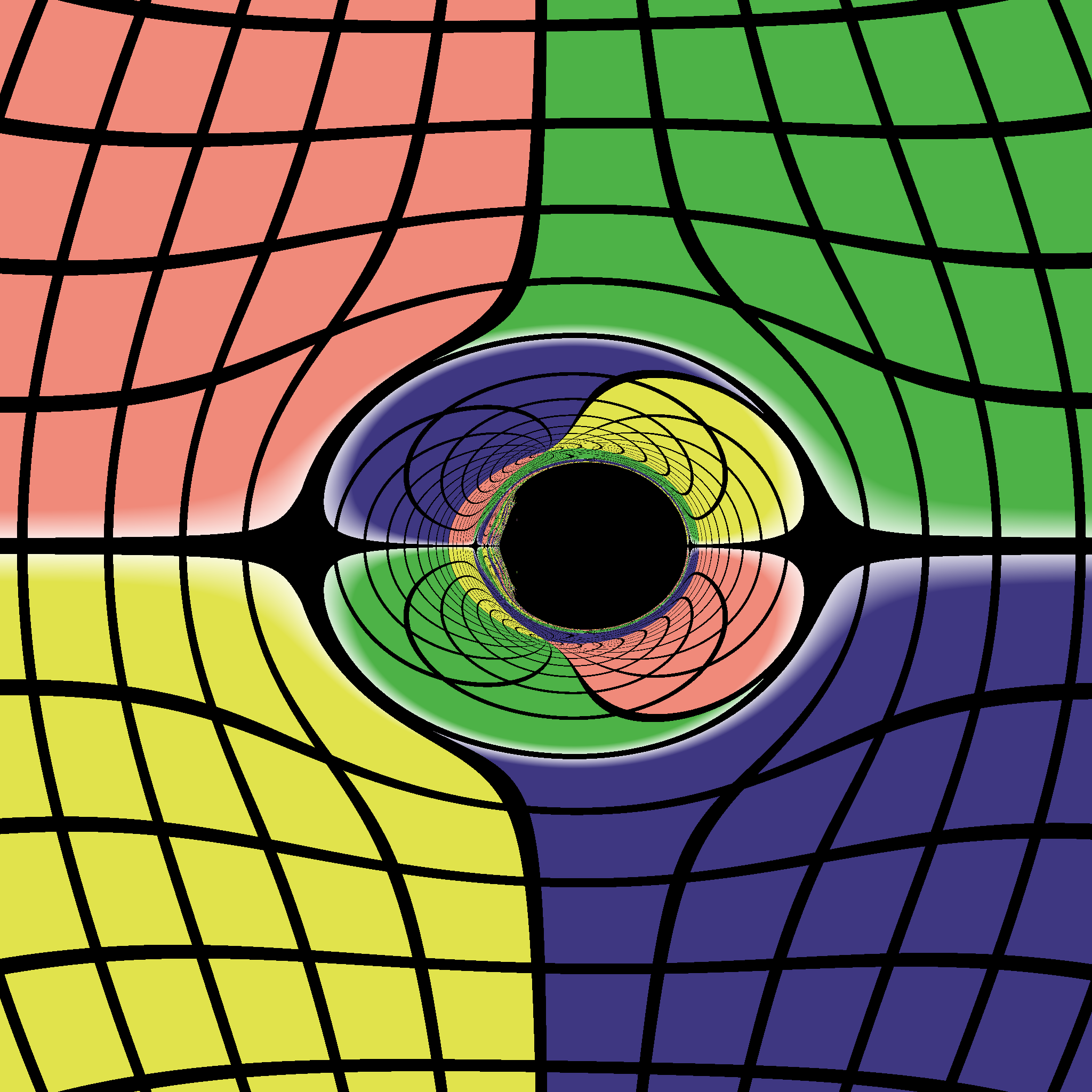}
  \end{minipage}
  }%

  \centering
  \caption{The images of the Kerr black hole in uniform magnetic fields with parameter $\Lambda=-0.01$. The inclination angle of the observer is fixed at $\theta_o=\pi/2$.}
  \label{variation_a}
\end{figure}

\begin{figure}[h!]

  \centering
  \includegraphics[width=3.4in]{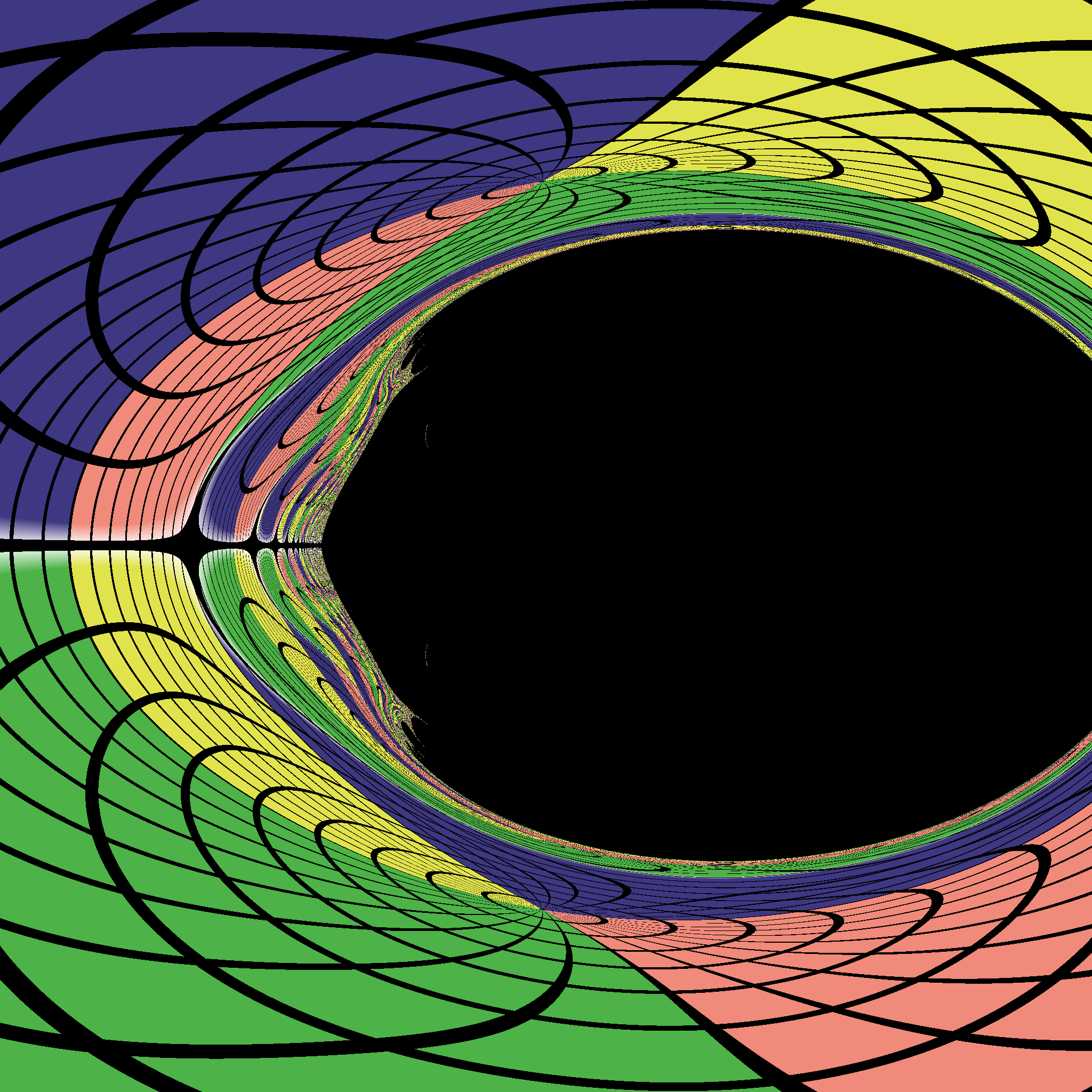}

  \caption{The fine structure of the photon hairs of (f) in Fig. \ref{variation_a}}
  \label{kerr}
\end{figure}

In Fig. \ref{variation_a}, we set $\Lambda=-0.01$ and $\theta_o=\frac{\pi}{2}$, and let the spin of the Kerr black hole $a$ go from $0.2$ to $0.999$. Obviously, when $a$ is small, the deformation of the shadow  is very small, the shadow looks very similar to the corresponding original Kerr black hole shadow. As the spin $a$ goes up, the left part of the shadow curve slowly shrinks, around the boundary, some photon ``hairs'' arise. In Fig. \ref{kerr}, we zoom in part of (f) with photon hairs in Fig. \ref{variation_a}. In particular, when the spin is very near extremality, the photon hair becomes apparent. In addition, one can obviously find a region of discontinuity at the top-left corner of the shadow curve, and symmetrically a similar region in the right corner. These observations are significantly different with the shadow of the Kerr black hole without magnetic field, in which the left boundary looks like a vertical line when the Kerr black hole is near extreme. These new features provide a potential way to verify whether the magnetic fields exist outside the horizon of a Kerr black hole if the EHT has enough capabilities to get a clear enough image of the black hole.

\begin{figure}[h!]
  \centering

  \subfigure[$\Lambda=-0.01$]{
  \begin{minipage}[t]{0.3\linewidth}
  \centering
  \includegraphics[width=1.5in]{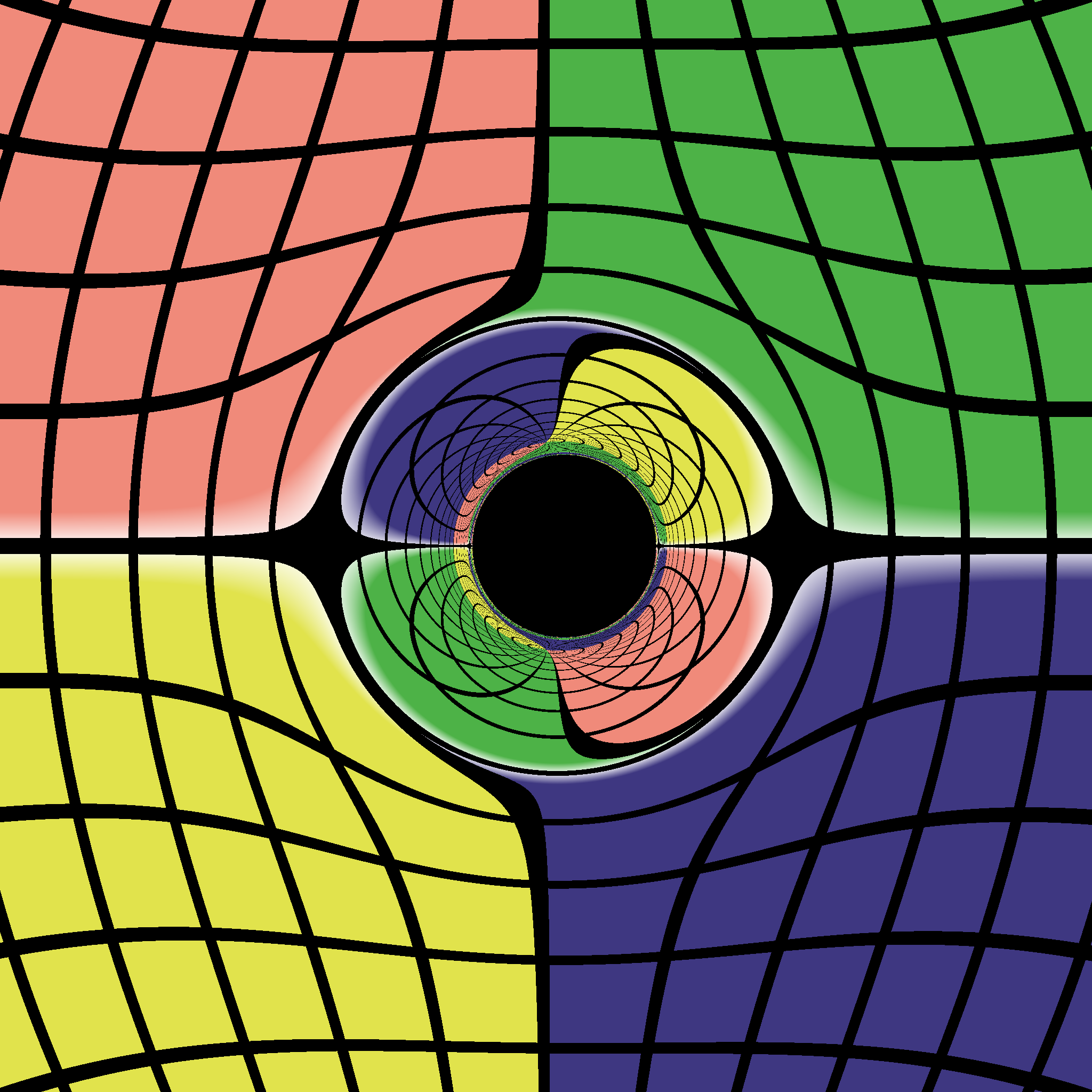}
  \end{minipage}%
  }%
  \subfigure[$\Lambda=-0.05$]{
  \begin{minipage}[t]{0.3\linewidth}
  \centering
  \includegraphics[width=1.5in]{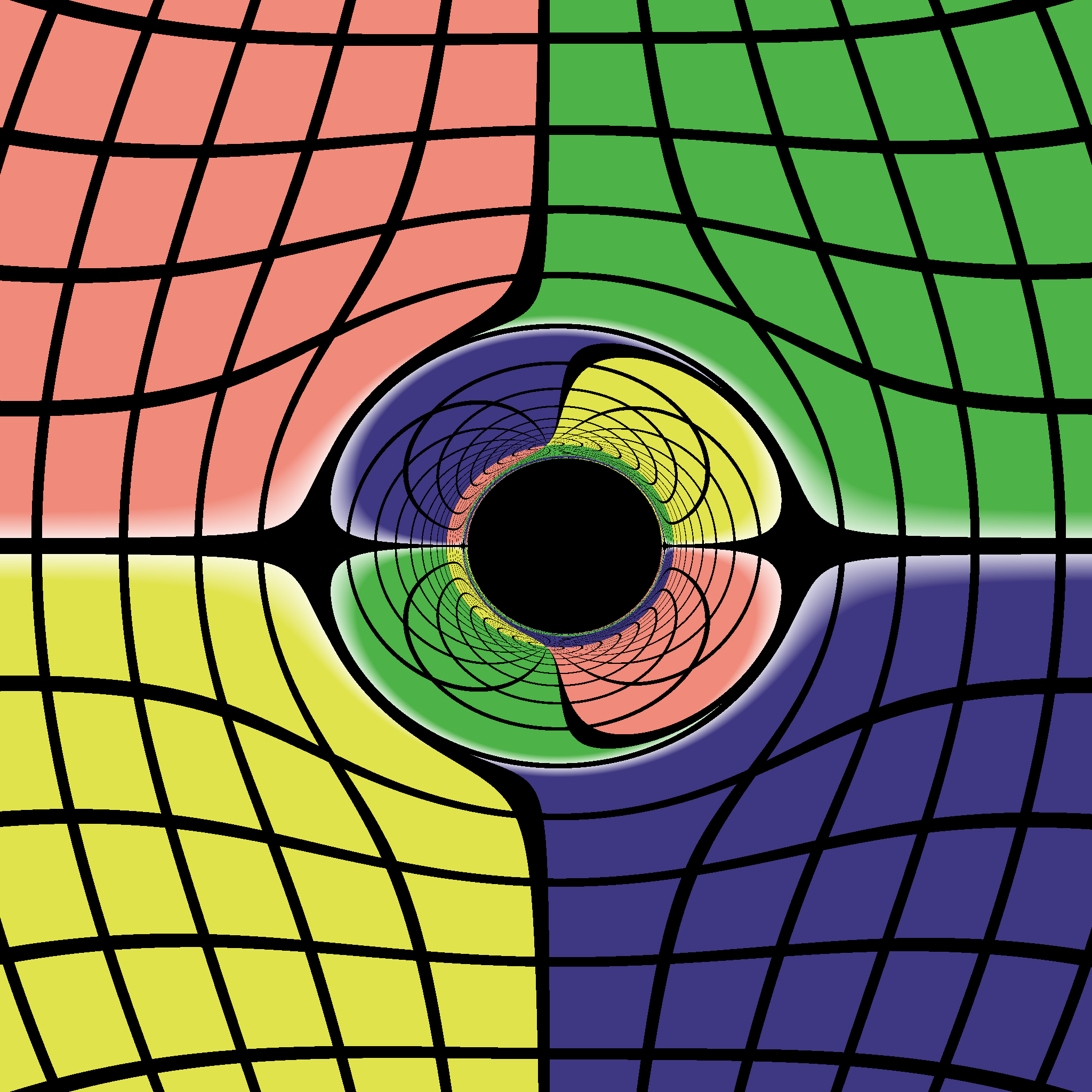}
  \end{minipage}%
  }%

  \subfigure[$\Lambda=-0.10$]{
  \begin{minipage}[t]{0.3\linewidth}
  \centering
  \includegraphics[width=1.5in]{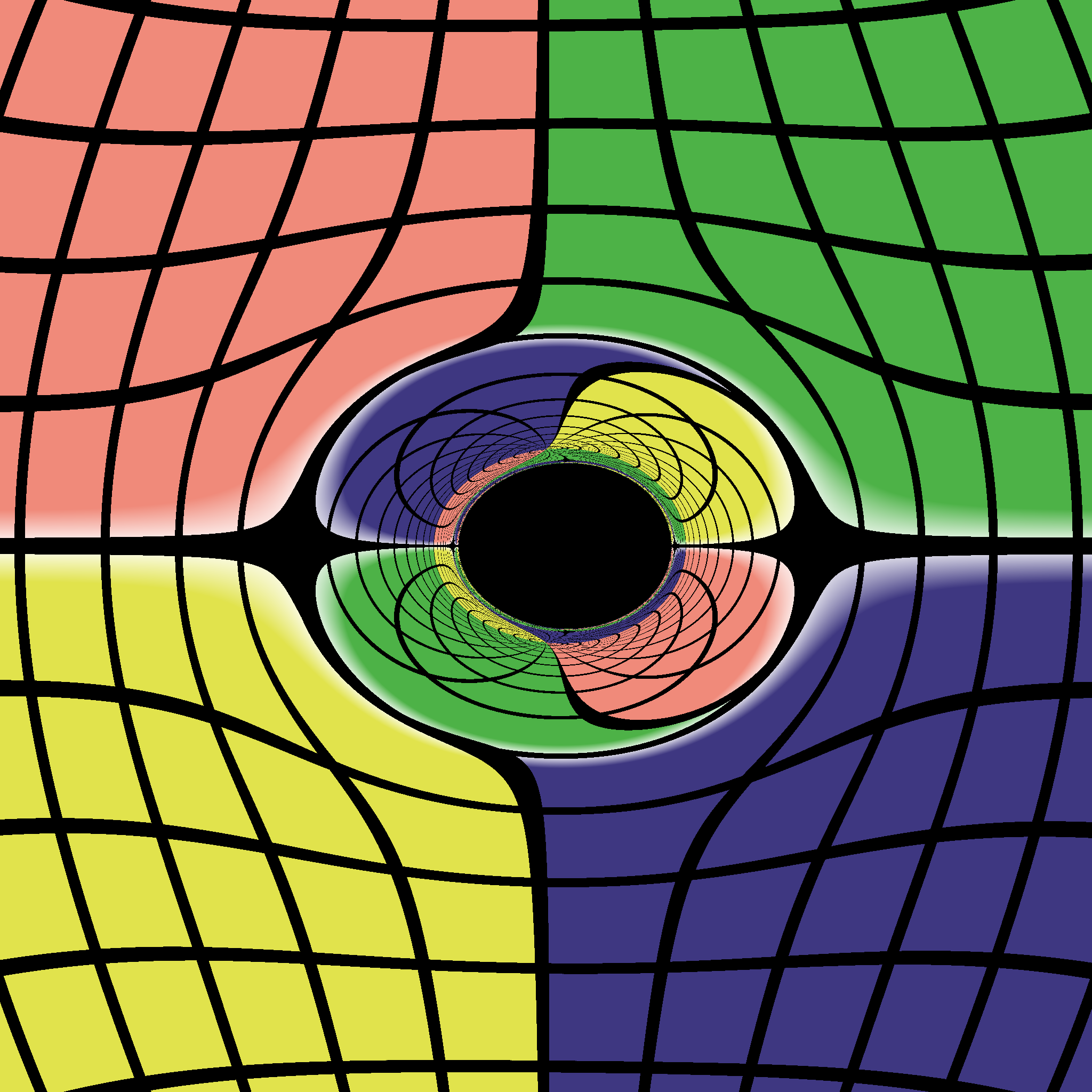}
  \end{minipage}
  }%
  \subfigure[$\Lambda=-0.15$]{
  \begin{minipage}[t]{0.3\linewidth}
  \centering
  \includegraphics[width=1.5in]{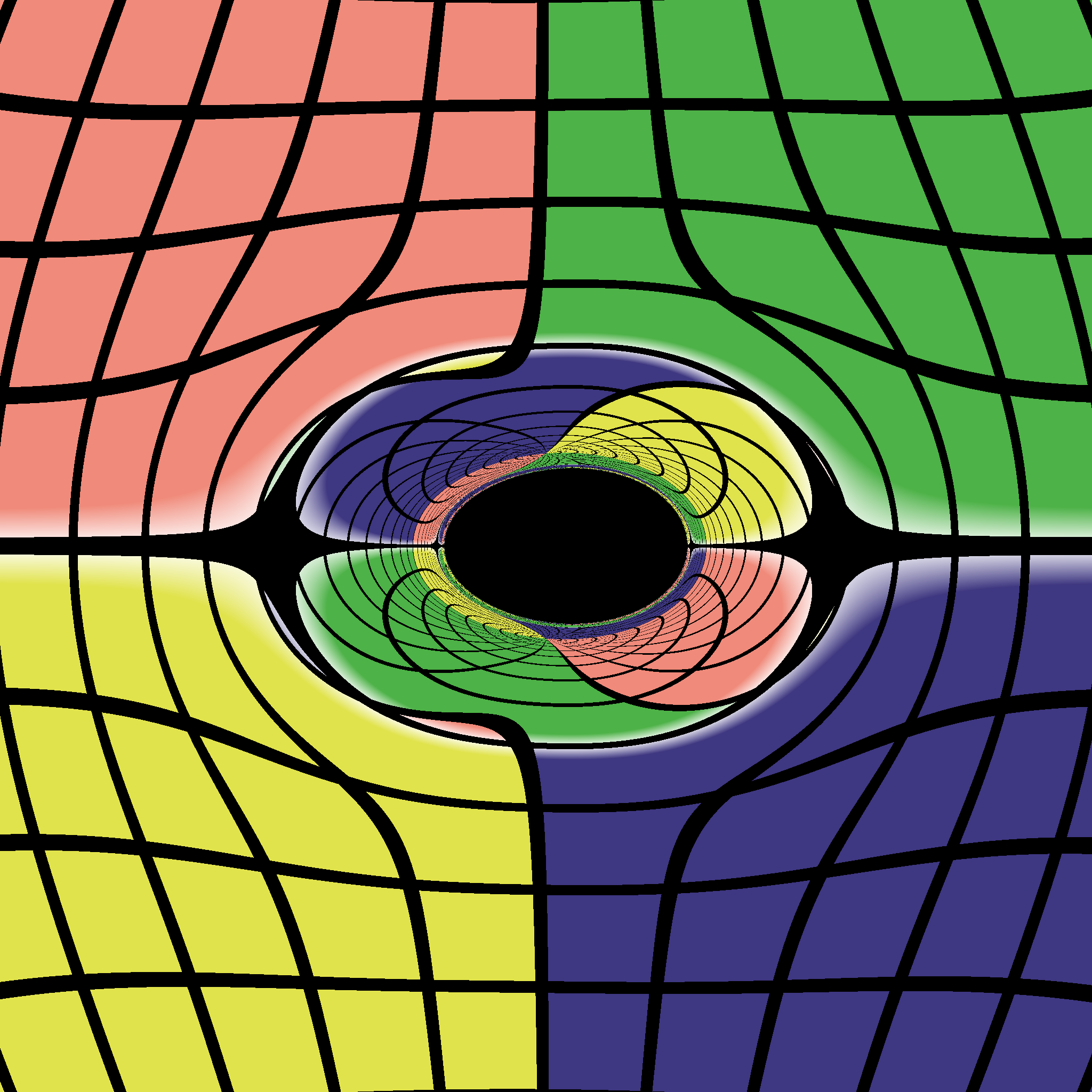}
  \end{minipage}
  }%
  \subfigure[$\Lambda=-0.20$]{
  \begin{minipage}[t]{0.3\linewidth}
  \centering
  \includegraphics[width=1.5in]{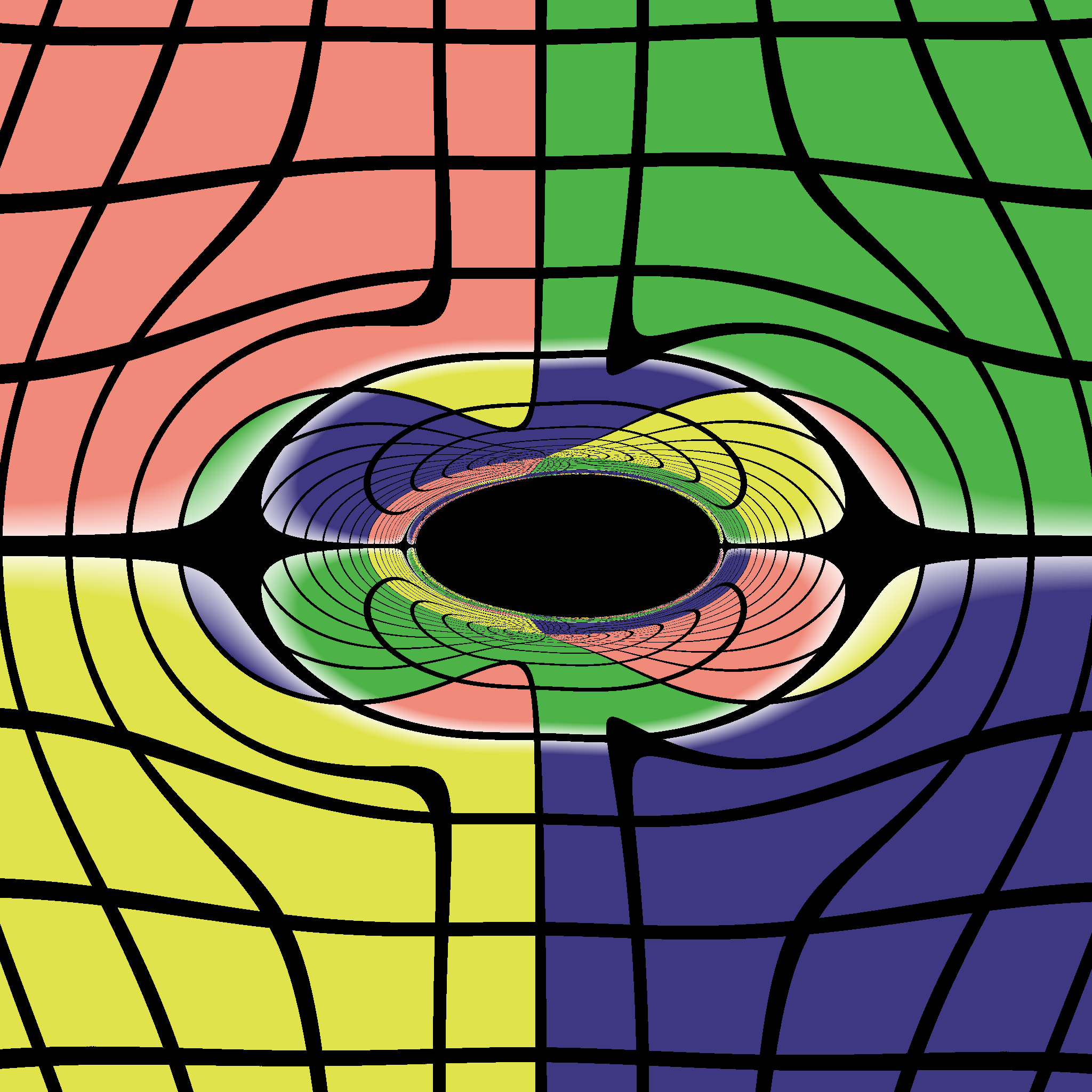}
  \end{minipage}
  }%

  \centering
  \caption{The images of the Kerr black hole in uniform magnetic fields. The inclination angle of the observer is fixed at $\theta_o=\pi/2$ and the spin is fixed at $a=0.5$.}
  \label{a=0.5}
\end{figure}

Next, let us move to the situation that the spin $a$ and the observational angle $\theta_o$ are fixed at $0.5$ and $\frac{\pi}{2}$, respectively. In Fig. \ref{a=0.5}, we can see the changes of the shadows  as the parameter $\Lambda$ vary from $-0.01$ to $-0.20$. The main feature is that the shadow curve is apparently stretched along the horizontal direction and squeezed along the vertical direction as the absolute value $|\Lambda|$ gets larger. In other words, when the strength of magnetic field gets stronger, the Kerr black hole shadow becomes more flattened. This is reminiscent of the  similar characteristic in the shadow of the Schwarzschild black hole bathed in a uniform magnetic. In fact, when $a$ is small, the frame dragging effect is negligible, so that many discussions on static black hole situations apply equally to the Kerr case, even though a little more complicated structures appear in the Kerr case, like the photon hairs in the neighborhood of the shadow curve. .

\begin{figure}[t]
  \centering

  \subfigure[$\Lambda=-0.01$]{
  \begin{minipage}[t]{0.3\linewidth}
  \centering
  \includegraphics[width=1.5in]{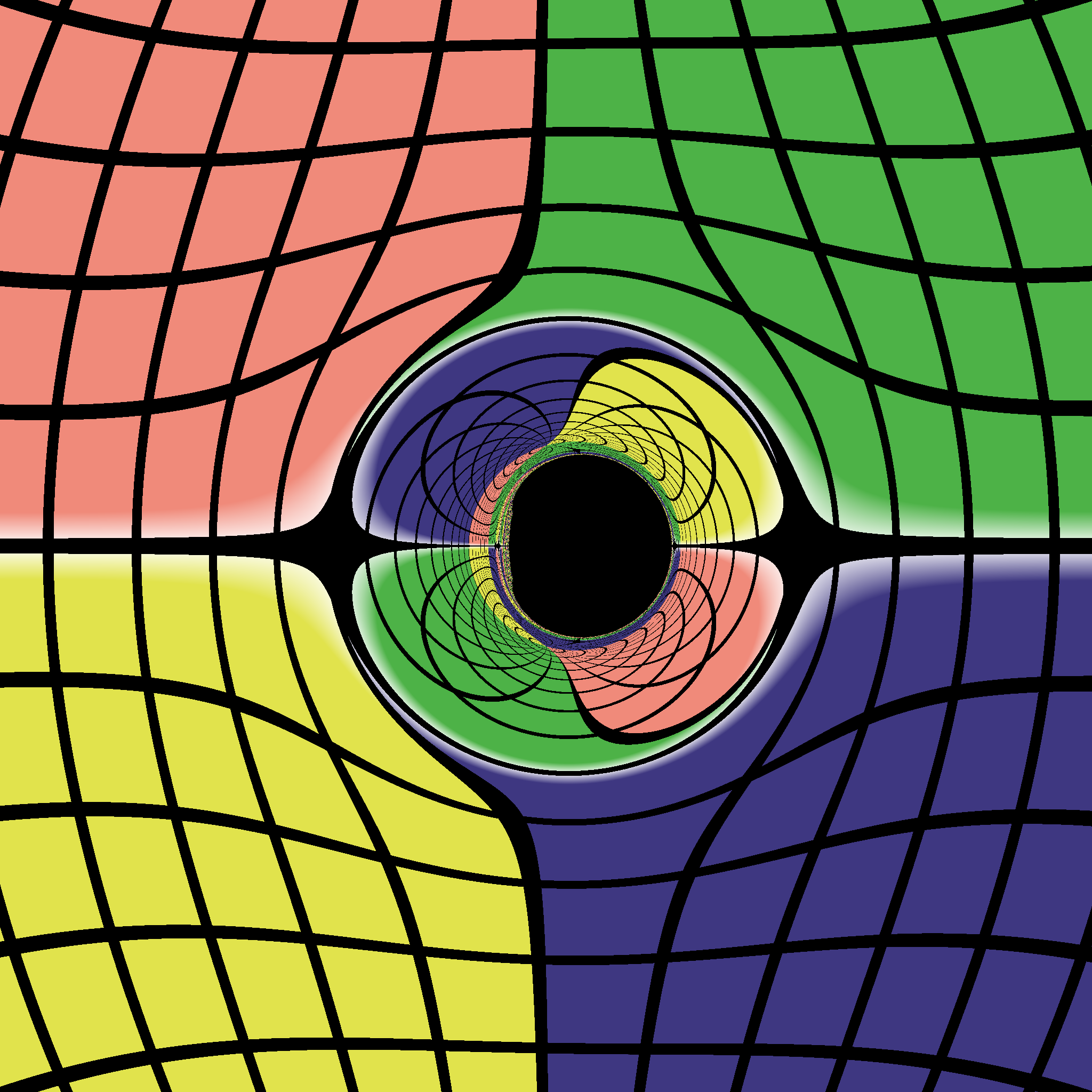}
  \end{minipage}%
  }%
  \subfigure[$\Lambda=-0.05$]{
  \begin{minipage}[t]{0.3\linewidth}
  \centering
  \includegraphics[width=1.5in]{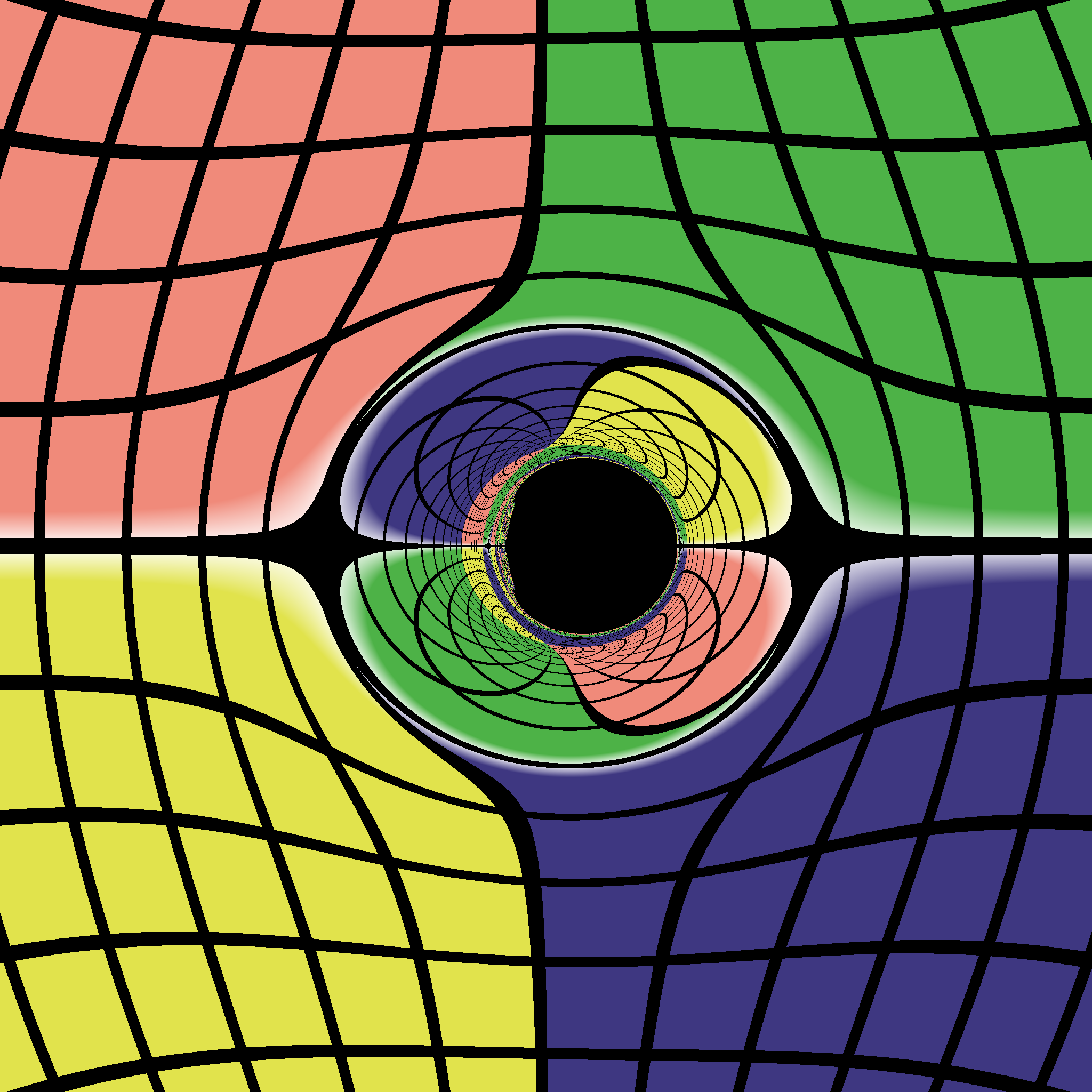}
  \end{minipage}%
  }%

  \subfigure[$\Lambda=-0.10$]{
  \begin{minipage}[t]{0.3\linewidth}
  \centering
  \includegraphics[width=1.5in]{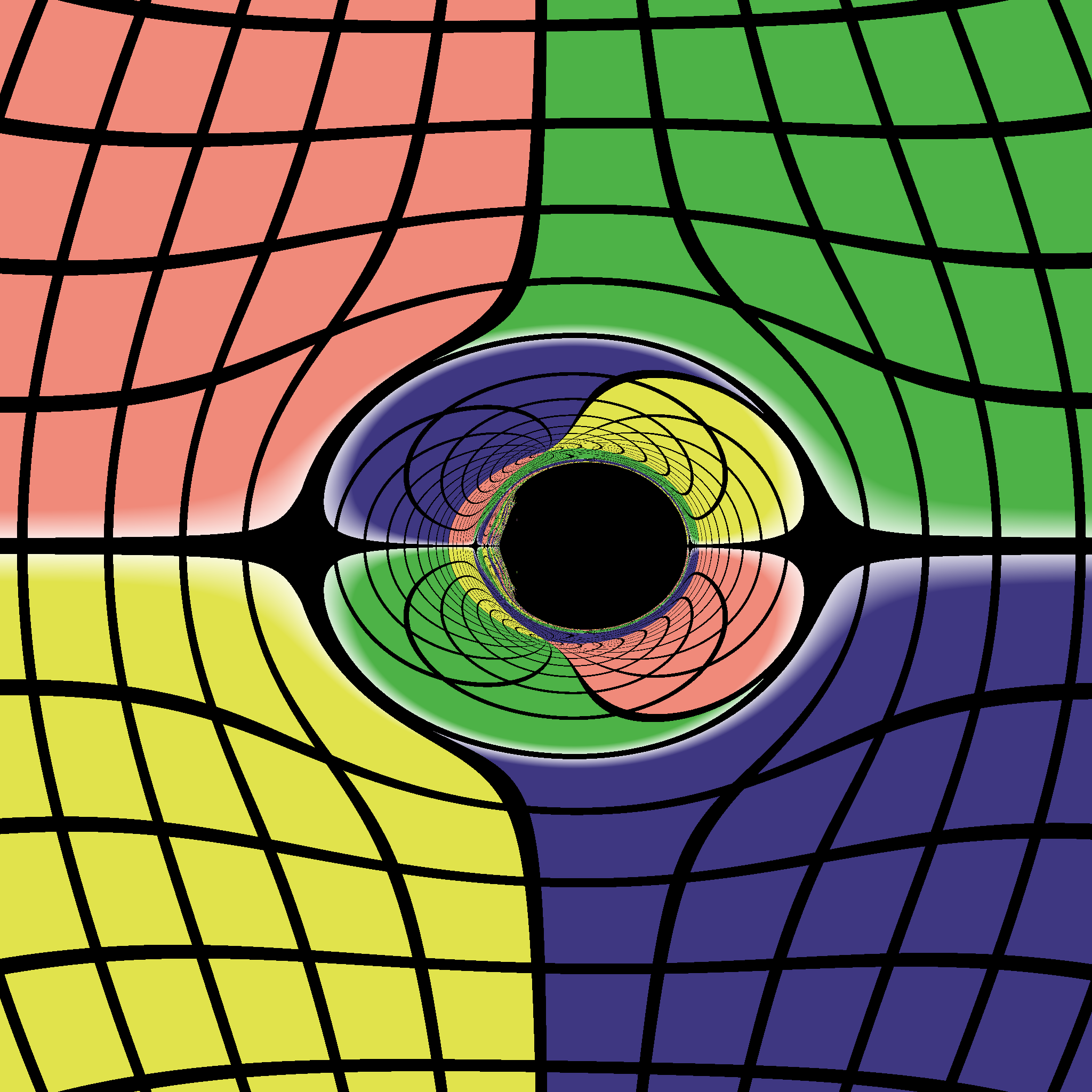}
  \end{minipage}
  }%
  \subfigure[$\Lambda=-0.15$]{
  \begin{minipage}[t]{0.3\linewidth}
  \centering
  \includegraphics[width=1.5in]{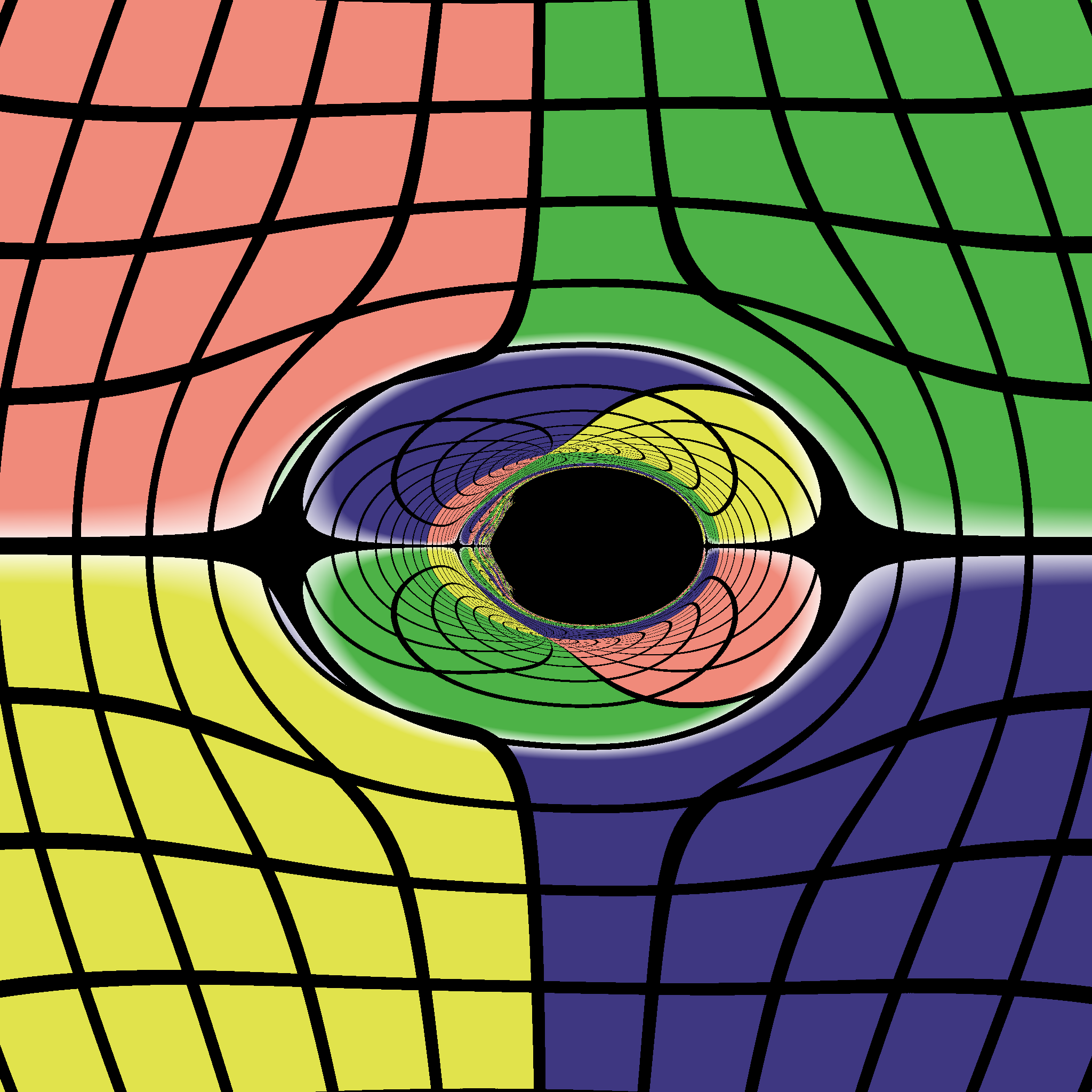}
  \end{minipage}
  }%
  \subfigure[$\Lambda=-0.20$]{
  \begin{minipage}[t]{0.3\linewidth}
  \centering
  \includegraphics[width=1.5in]{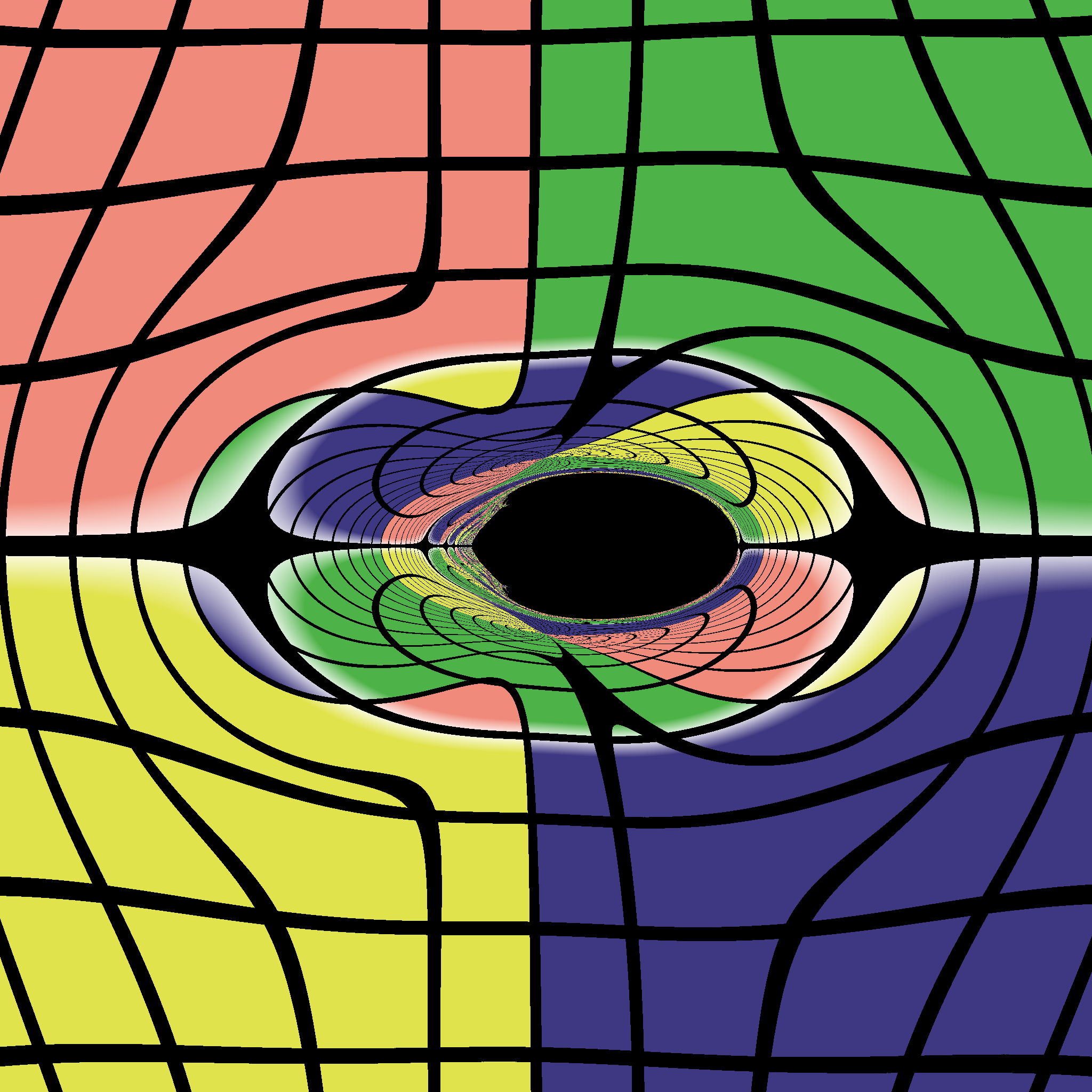}
  \end{minipage}
  }%

  \centering
  \caption{The images of the Kerr black hole in uniform magnetic fields. The inclination angle of the observer is fixed at $\theta_o=\pi/2$ and the spin is fixed at $a=0.999$.}
  \label{a=0.999}
\end{figure}

Now, let us look at a near-extremal Kerr black hole with $a=0.999$. We again fix the inclination angle of the observer at $\theta_o=\pi/2$ and let the parameter $\Lambda$ go from $-0.01$ to $-0.20$. From Fig. \ref{a=0.999},  we can clearly see that the photon hairs become very obvious at $a=0.999$ and, in addition to that, the shape of the shadow curve becomes more flattened as the parameter $|\Lambda|$ gets bigger. As shown in \cite{Cunha:2016bjh, Wang:2018eui, Chen:2016tmr, Wang:2016wcj, Wang:2021ara, Lima:2021cgb}, for non-integrable geodesic equations, the photons may travel in chaotic motions. In particular, in \cite{Lima:2021cgb, Wang:2021ara}, the authors investigated the shadows of the black hole immersed in the Melvin magnetic field, and observed the chaotic motions of the photons. Furthermore, in \cite{Wang:2021ara}, the authors found the photon hair appearing at the left of the shadow curve as well, which is very similar to our findings. However the photon hairs found in our work have a bit more complicated structure: they extend two small ``hairs" like a claw into the shadow region in a strange way, around the positions where the shadow curve is discontinuous. The appearance of the photon hairs  is certainly an important feature that happens in a rotational black hole spacetime with a magnetic field.  And different features in the photon hairs may help us to distinguish  a uniform magnetic field from a Melvin magnetic field.

\begin{figure}[t!]
  \centering

  \subfigure[$\theta_0=0.1\pi$]{
  \begin{minipage}[t]{0.3\linewidth}
  \centering
  \includegraphics[width=1.5in]{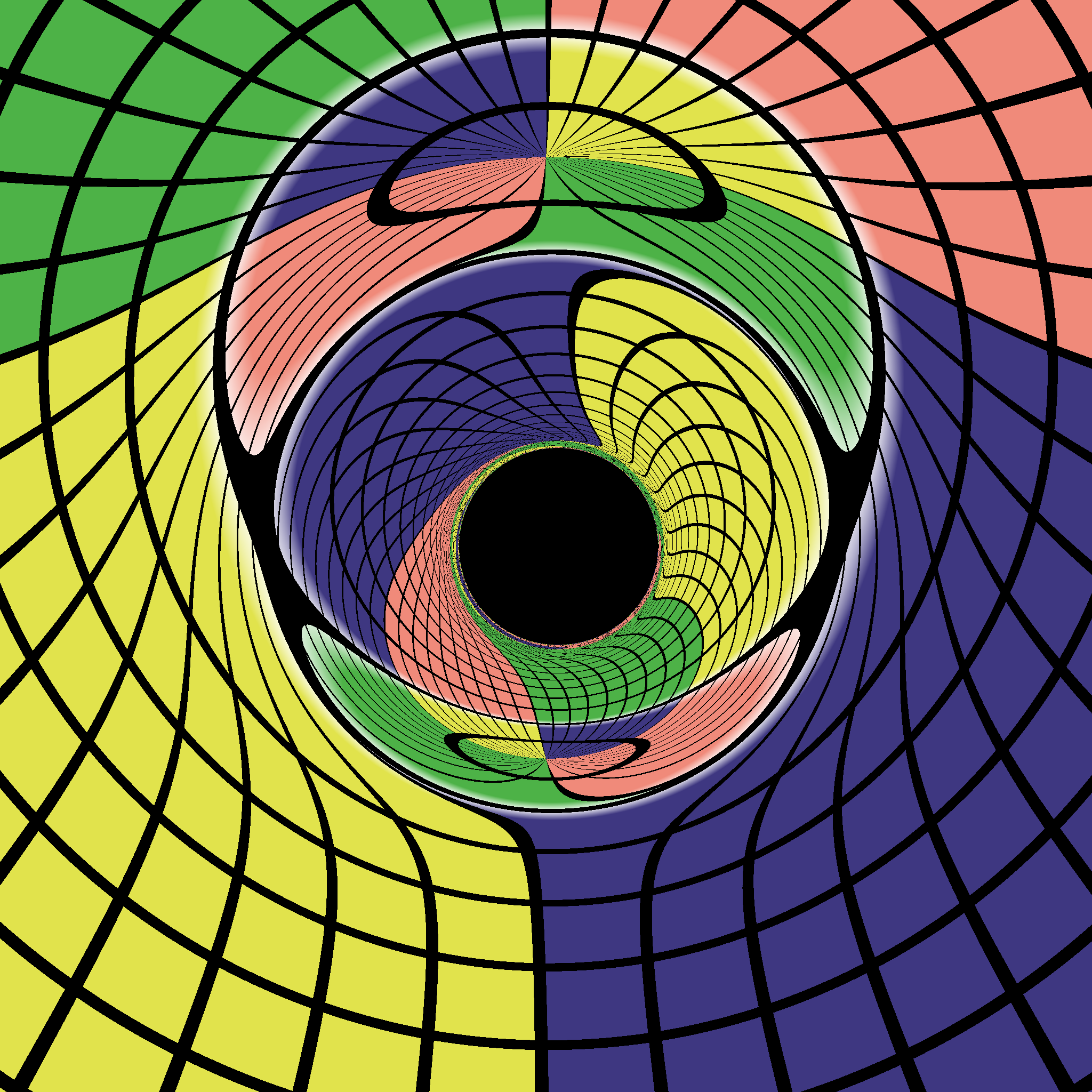}
  \end{minipage}%
  }%
  \subfigure[$\theta_0=0.3\pi$]{
  \begin{minipage}[t]{0.3\linewidth}
  \centering
  \includegraphics[width=1.5in]{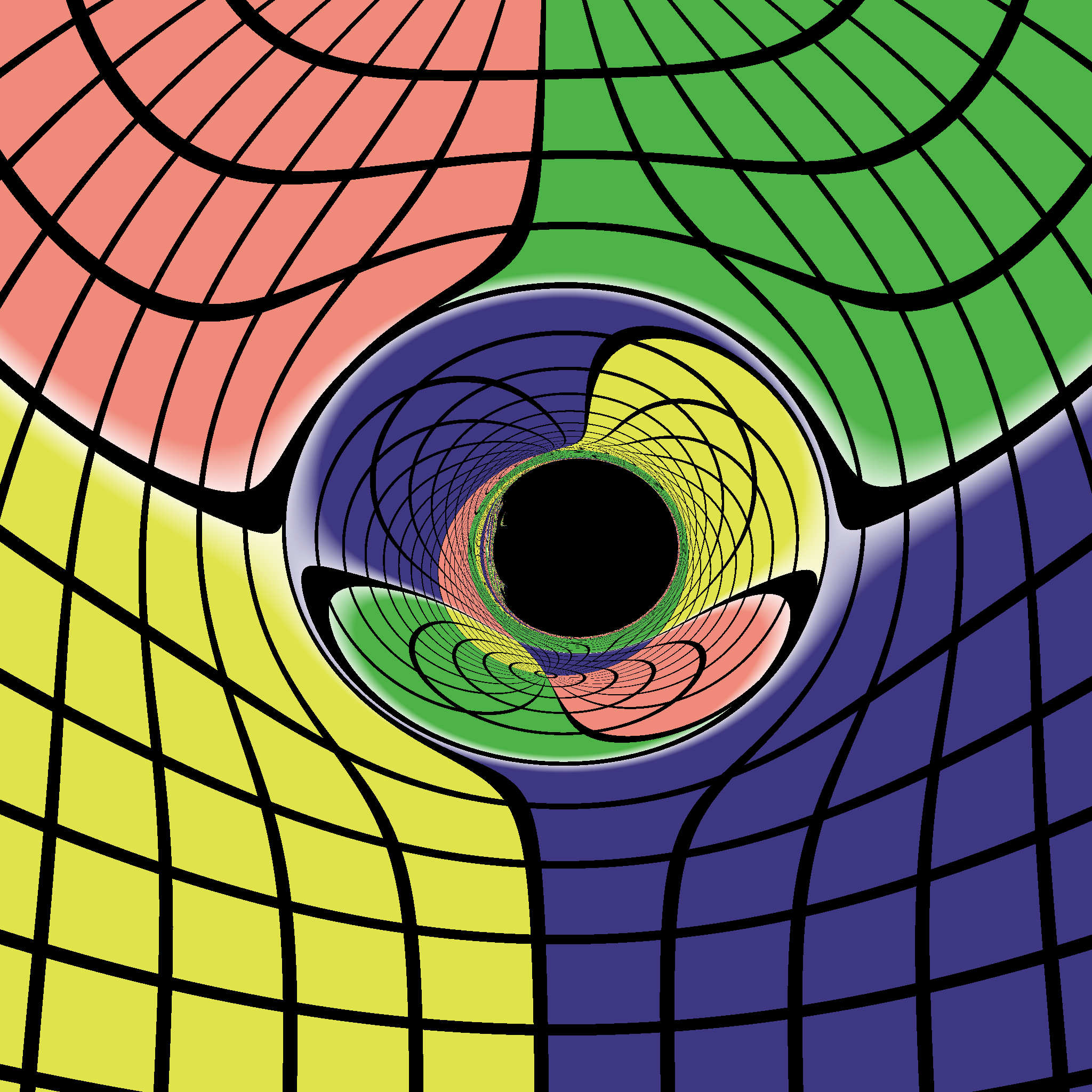}
  \end{minipage}%
  }%
    \subfigure[$\theta_0=0.5\pi$]{
  \begin{minipage}[t]{0.3\linewidth}
  \centering
  \includegraphics[width=1.5in]{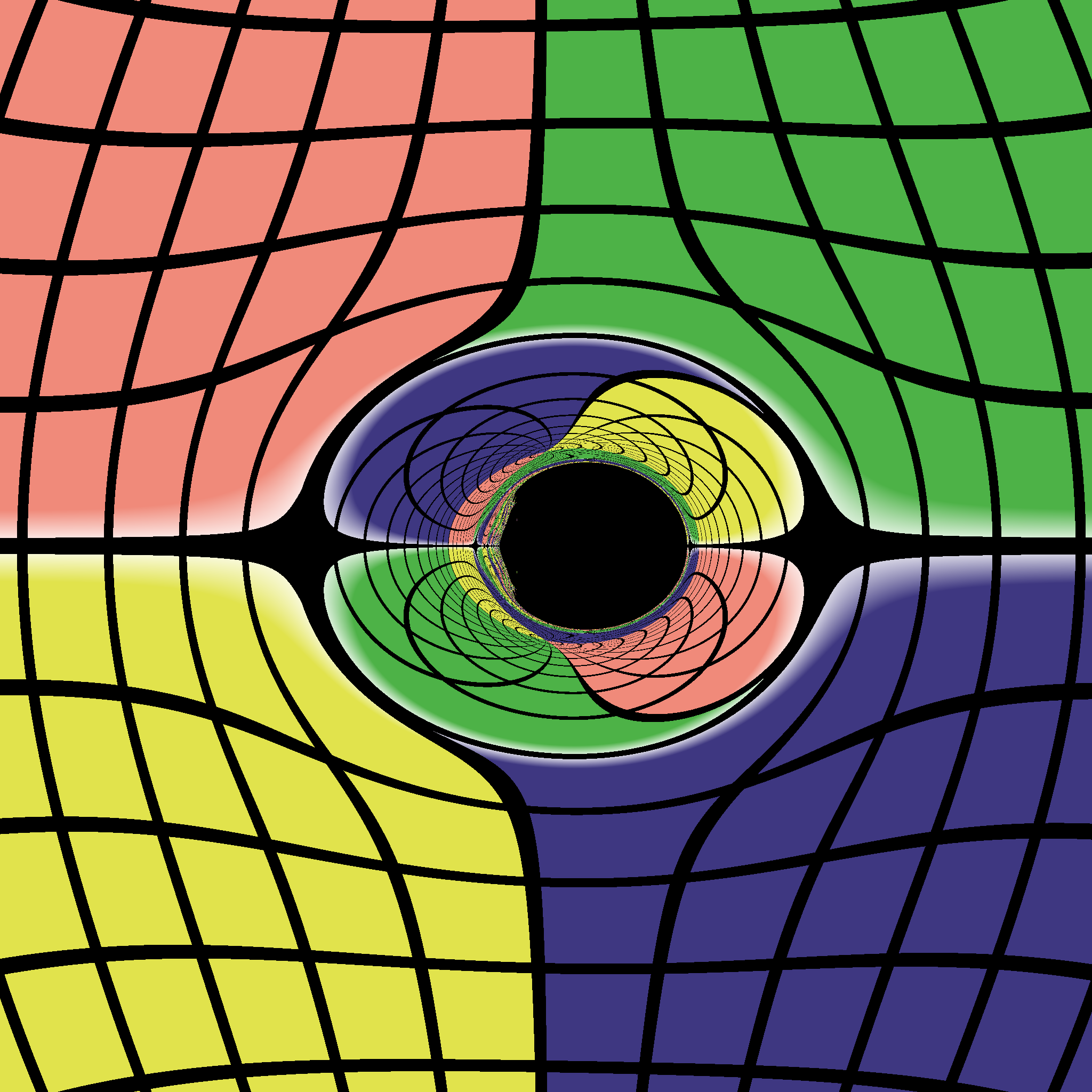}
  \end{minipage}%
  }%
  \centering
  \caption{The images of the Kerr black hole in uniform magnetic fields with $a=0.999$ and $\Lambda=-0.1$.}
  \label{kerr1}
\end{figure}

Moreover, it is also interesting to photograph the Kerr black hole from different inclination angles. Note that as the effective metric still has $\mathbb{Z}_2$ symmetry, we just have to consider $0\le\theta_o\le\pi/2$. Three examples are shown in Fig. \ref{kerr1}. We can see that the shadow is almost a perfect circle when $\theta_o$ is small, say $0.1\pi$, because of the existence of Killing vector field $\partial_{\phi}$. As $\theta_o$ increase, it produces some irregular structures. In the picture (b), the upper small ``hair" is separated and appears in the interior of the shadow. Finally, the irregular structure becomes symmetric along the horizontal line at $\theta_o=\pi/2$ due to the $\mathbb{Z}_2$ symmetry of the system.

\subsection{Quantitative description of the deformation}\label{qdd}

\begin{figure}[h!]

  \centering
  \includegraphics[width=3.4in]{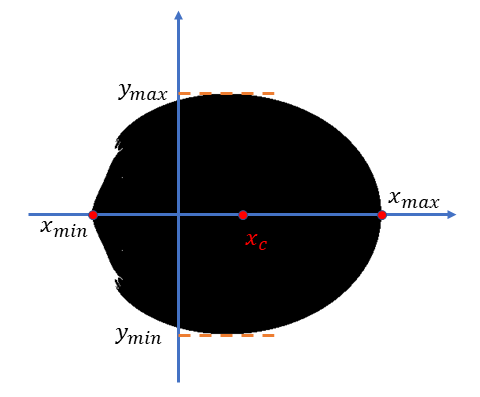}

  \caption{A diagram of the Kerr black hole shadow curve, deformed by the QED effect from a uniform magnetic field around a black hole. The camera is located  on the equatorial plane.}
  \label{showing}
\end{figure}

In this subsection, we would like to use characteristic parameters to quantify the deformation of the  shadow of a Kerr black hole, due to a uniform magnetic field and its induced QED effect. Following \cite{Johannsen:2013vgc, Cunha:2015yba}, we introduce six parameters, $\{D_c,D_x,D_y,\bar{r},\sigma_r,\sigma_{Kerr}\}$ to study  the deviation driven by the QED effect quantitatively. To better interpret the meaning of these parameters, we present a diagram in Fig. \ref{showing}, in which we place the shadow curve in the Cartesian coordinates $(x,y)$ with the origin being the crosspoint of the boundary of four parts with different colors in the celestial sphere. And $x_\text{min}$ and $x_\text{max}$ are the smallest and the largest values that can be taken from the points of the shadow's edge on the horizontal axis. Similarly, we get $y_\text{min}$ and $y_\text{max}$, on the vertical axes. Note the $\mathbb{Z}_2$ symmetry of the effective metric, we have $y_{\text{min}}+y_\text{max}=0$ for the cameras sitting on the equatorial plane.

Then, we can easily introduce the definitions of other  parameters. The center of the shadow is defined as
\bea
x_c\equiv\frac{x_\text{max}+x_\text{min}}{2}
\eea
on the horizontal axes. And $y_c$ is defined similarly which is exactly zero due to the $Z_2$ symmetry. Thus the first parameter $D_c\equiv|x_c|$ measures the distance from the center of the shadow to the coordinate origin. And $D_x\equiv x_{\text{max}}-x_\text{min}$ and $D_y\equiv y_\text{max}-y_\text{min}$ are the width and the height of the shadow respectively.

\begin{table}[h!]
\centering
\caption{Quantitative parameters characterizing shadows, with $a=0.999$ and $\theta_0=\pi/2$. }
\begin{tabular}{ccccccccc}
\hline
Parameters & $\Lambda$ & $D_c$ & $D_x$ & $D_y$ & $\bar{r}$ & $\sigma_r$ & $\sigma_r/\bar{r}$ & $\sigma_{K}$ \\
\hline
Values & -0.01 &  82.5 & 306 & 342 & 163.6 &  8.69 & 0.0531 & 0.0075   \\
       & -0.03 &  83.5 & 314 & 336 & 163.8 &  6.95 & 0.0424 & 0.0210   \\
       & -0.05 &  84.5 & 322 & 330 & 164.2 &  5.35 & 0.0326 & 0.0368   \\
       & -0.07 &  86.0 & 333 & 324 & 164.8 &  4.85 & 0.0294 & 0.0546   \\
       & -0.09 &  87.0 & 345 & 316 & 165.7 &  5.89 & 0.0355 & 0.0756   \\
       & -0.11 &  89.5 & 360 & 310 & 166.9 &  8.69 & 0.0521 & 0.0992   \\
       & -0.13 &  91.5 & 378 & 304 & 168.5 & 12.42 & 0.0737 & 0.1282   \\
       & -0.15 &  94.5 & 400 & 296 & 170.7 & 17.20 & 0.1007 & 0.1631   \\
       & -0.16 &  96.5 & 414 & 292 & 172.1 & 20.06 & 0.1166 & 0.1838   \\
       & -0.17 &  99.0 & 431 & 288 & 173.6 & 23.97 & 0.1380 & 0.2098   \\
       & -0.18 & 102.0 & 449 & 284 & 175.5 & 27.06 & 0.1542 & 0.2340   \\
       & -0.19 & 105.0 & 471 & 280 & 177.6 & 31.47 & 0.1772 & 0.2659   \\
       & -0.20 & 109.5 & 498 & 274 & 180.0 & 20.06 & 0.2039 & 0.3033   \\
\hline

\end{tabular}
\label{table}
\end{table}

From another perspective, we can define the polar coordinates $(r,\theta)$ with the origin at the center of the shadow, which means $r=[(x-x_{c})^2+y^2]$. Then the average radius is defined as
\bea
\bar{r}\equiv \int^{2\pi}_{0} r(\theta) d \theta / 2\pi\,,
\eea
and the deviation from the sphericity is defined to be 
\bea
\sigma_{r}\equiv \left\{\int^{2\pi}_{0} [r(\theta)-\bar{r}]^2 d \theta /2\pi \right\}^{1/2}\,.
\eea
 At last, the deviation from a comparable Kerr BH ($\Lambda=0$) is
\be
\sigma_{K}\equiv \sqrt{\frac{1}{2\pi} \int^{2\pi}_{0} \left(\frac{r(\theta)-r_{Kerr}(\theta)}{r_{Kerr}(\theta)} \right)^2 d \theta}.
\ee

\begin{figure}[h!]
  \centering

  \subfigure[$D_y/D_x$]{
  \begin{minipage}[t]{0.5\linewidth}
  \centering
  \includegraphics[width=3in]{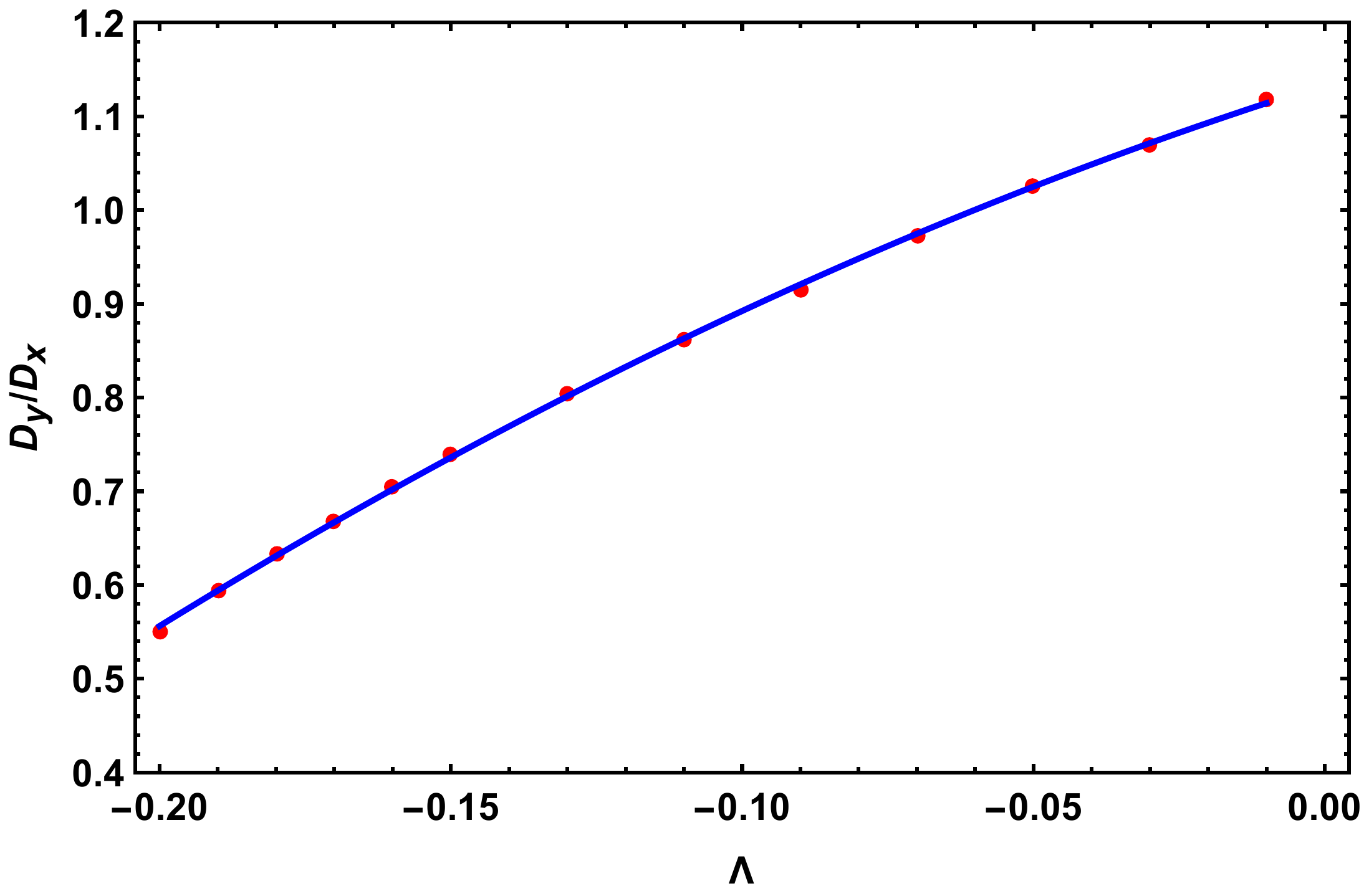}
  \end{minipage}%
  }%
  \subfigure[$\bar{r}$]{
  \begin{minipage}[t]{0.5\linewidth}
  \centering
  \includegraphics[width=3in]{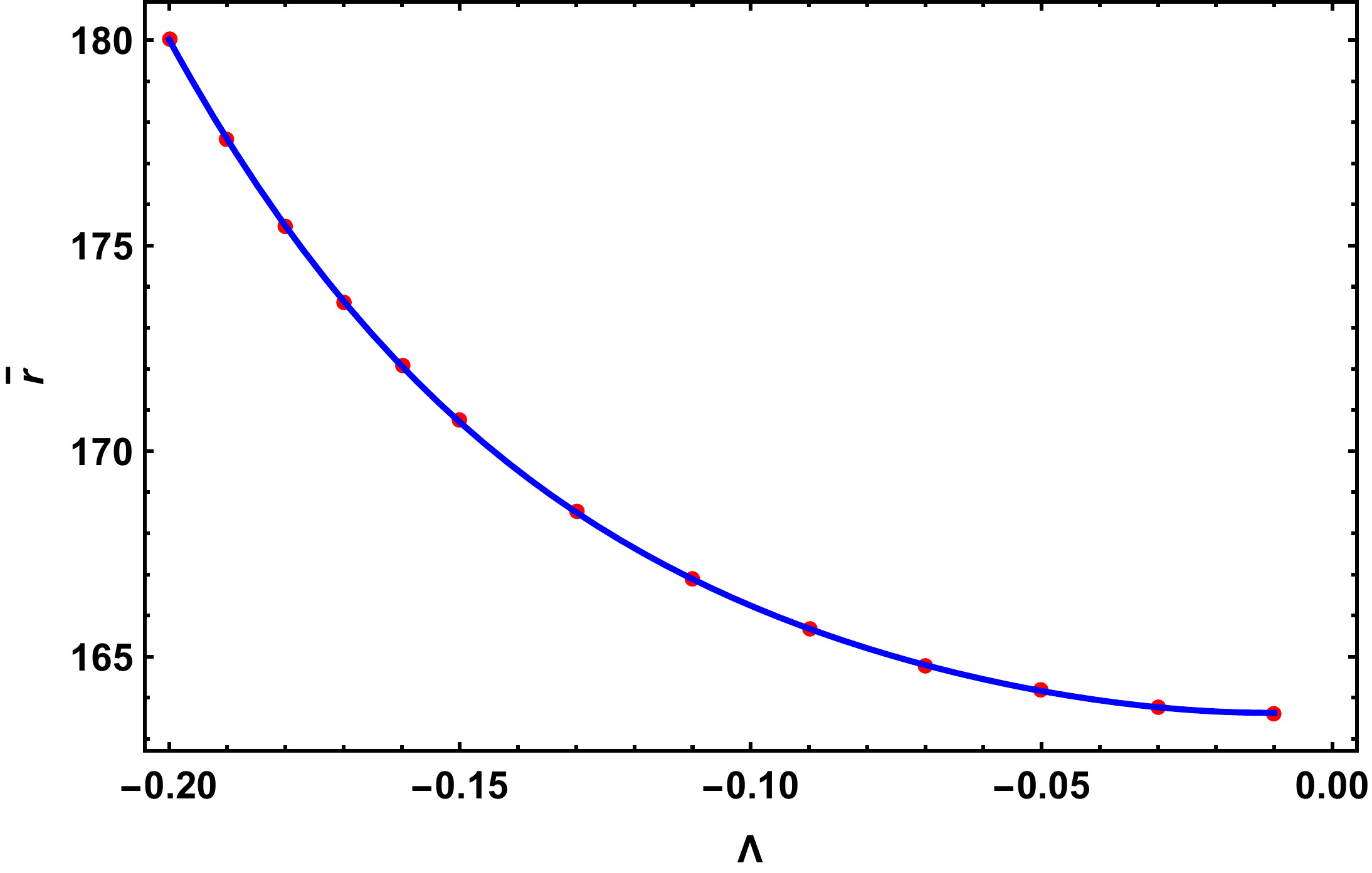}
  \end{minipage}%
  }%

    \subfigure[$\sigma_r/\bar{r}$]{
  \begin{minipage}[t]{0.5\linewidth}
  \centering
  \includegraphics[width=3in]{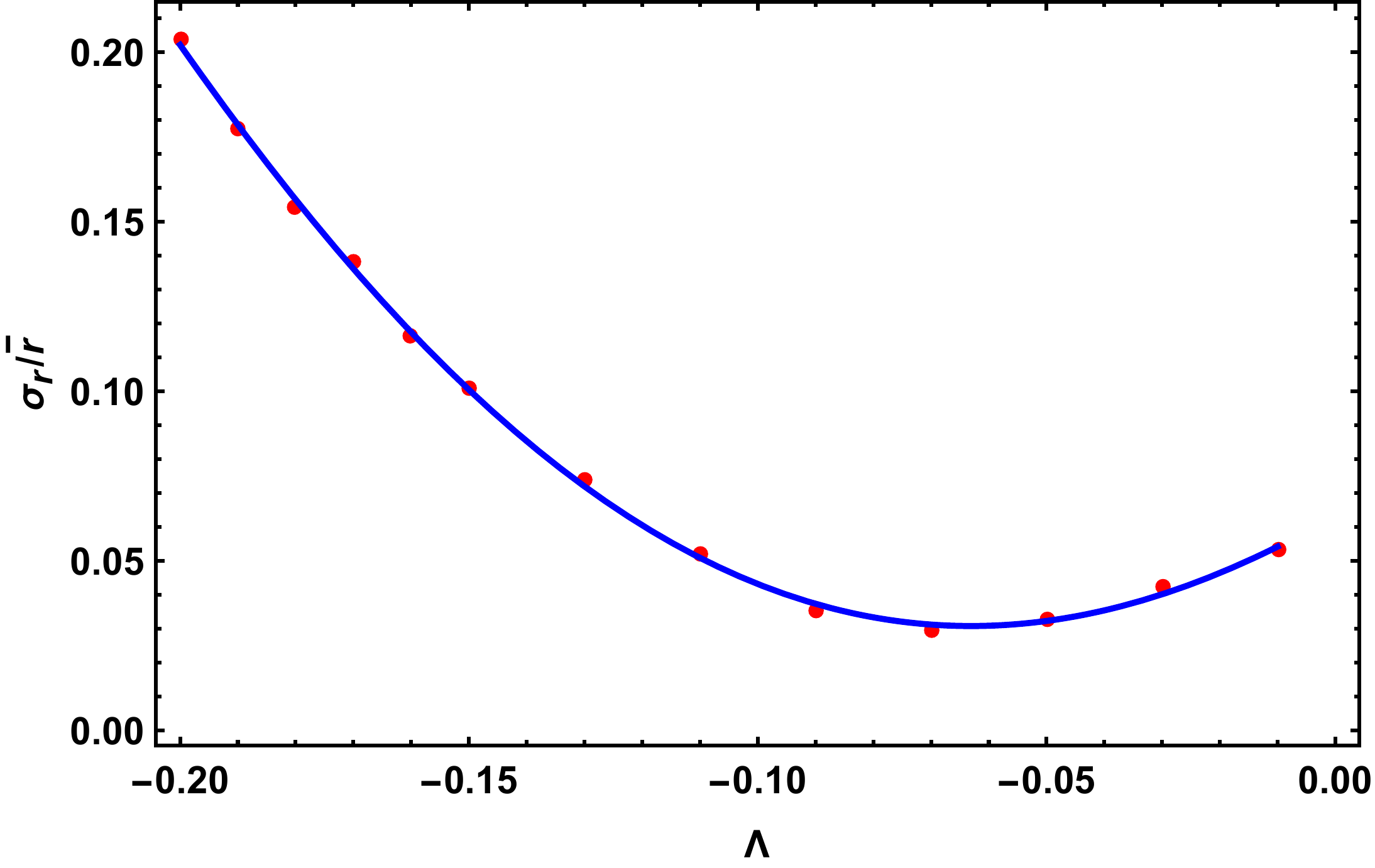}
  \end{minipage}%
  }%
  \subfigure[$\sigma_K$]{
  \begin{minipage}[t]{0.5\linewidth}
  \centering
  \includegraphics[width=3in]{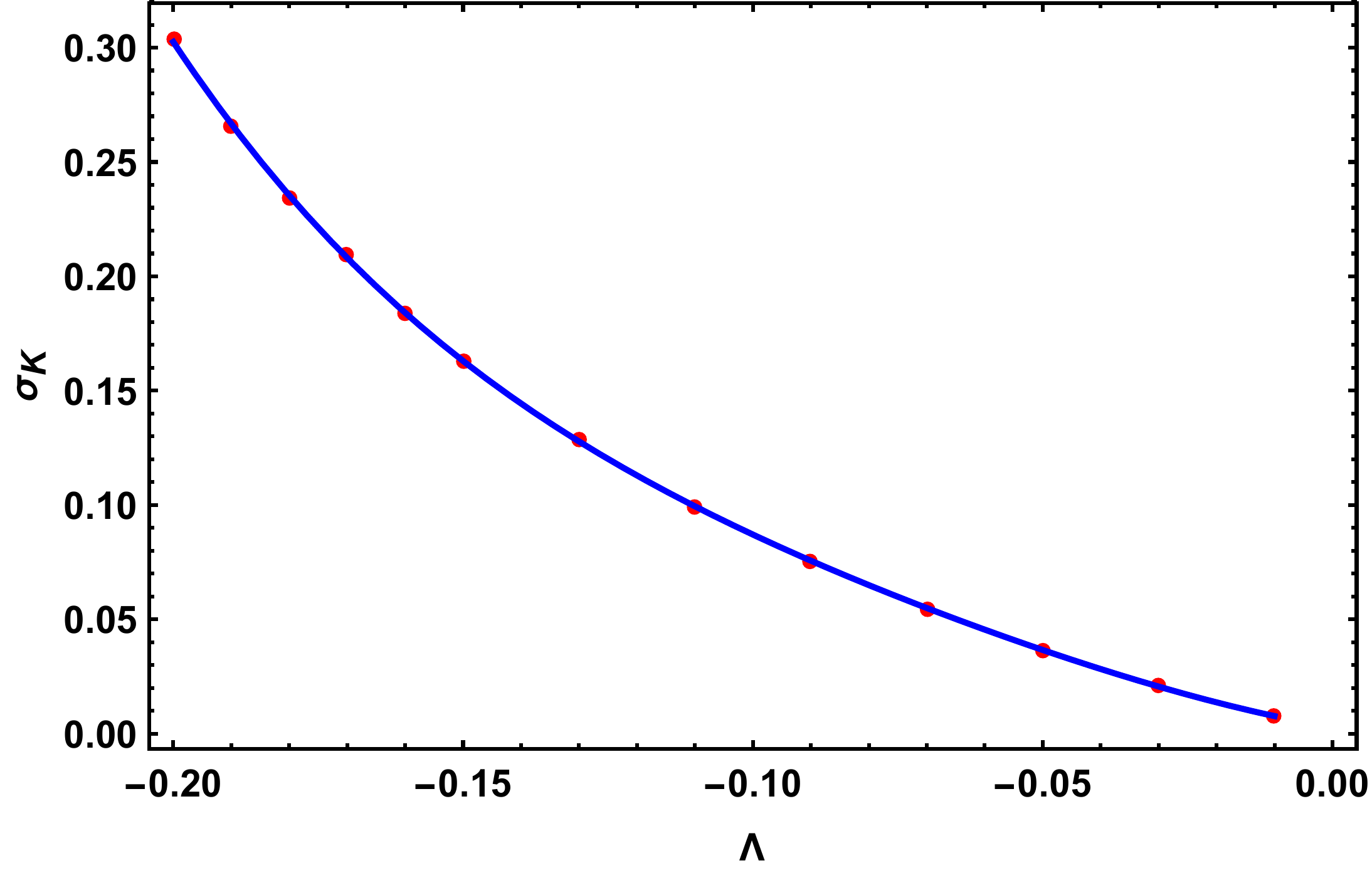}
  \end{minipage}%
  }%
  \centering
  \caption{The variations of different parameters with respect to $\Lambda$. Here we fix $a=0.999$ and $\theta_0=\pi/2$.}
  \label{para}
\end{figure}

In Table. \ref{table}, we show these parameters describing the shadow when the spin $a=0.999$ and the observational angle $\theta_0=\pi/2$. In Fig. (\ref{para}), we illustrate the variations of $D_y/D_x$, $\bar{r}$, $\sigma/\bar{r}$ and $\sigma_K$ with respect to $\Lambda$.  Apparently, a stronger magnetic field leads to the increase of $D_x$ and the decrease of $D_y$, which can be seen easily in Table. \ref{table}. Thus, in Fig. \ref{para}, we can see that the quantity $D_y/D_x$, which can somehow be regarded as oblateness, is an increasing function for $-0.2\le\Lambda\le-0.01$. A similar observation also
appears in the static black hole case which have been discussed in detail in \cite{Hu:2020usx}. An interesting thing is that when $a\ge0$, we find that the deviation $D_c$ becomes larger as the strength of the magnetic field increases. However, we are not sure if this finding has some useful implications in astronomical observations.

In addition, from the data of the average radius and the deviation from the sphericity in Table. \ref{table}, we can see that both of them go up when $|\Lambda|$ goes from $0.01$ to $0.20$, which means that an increasing uniform magnetic field would enlarge the circumference of the shadow. However, from Table. \ref{table}, as well as Fig. \ref{para}, we find that the deviation from the sphericity $\sigma_r$ and the deviation per average radius $\sigma_r/\bar{r}$ are not monotonic functions of $\Lambda$, instead, they start to go down and then go up when the strength of magnetic field grows from weak to strong.

Finally, let us turn to the deviation from a comparable Kerr BH ($\Lambda=0$), that is, $\sigma_K$. Both from Table. \ref{table} and Fig. \ref{para}, the derivation is found to increase with the strength of uniform magnetic fields. Furthermore, the rate of increase of the derivation also gets larger and larger as the magnetic field increases. These facts reveal that, for the Kerr black holes immersed in a uniform magnetic field with considerable strength,  the shape of the shadow would be significantly different from that of the vacuum case, due to the QED effect. This result provides a potential way to measure the magnetic field outside a real black hole in our Universe.

\section{Summary}\label{summary}

In this paper, we studied the QED effects on the shadows of the Kerr black holes immersed in magnetic fields. This  generalized our previous study on static black holes \cite{Hu:2020usx} to stationary rotating black holes. In the Kerr black hole case,  the effective metric involving the QED effects of the photons become much more complicated (see the appendix \ref{appendixA}). Thus, we only focused on the uniform magnetic fields outside a Kerr black hole in this work. We mainly investigated the influence of the uniform magnetic fields on the shadows of Kerr black holes  by employing the numerical backward ray-tracing method. We introduced a new dimensionless parameter $\Lambda=\lambda B^2$ to characterize the strength of the magnetic field. To guarantee the causality of the effective metric, we confined the parameter $\Lambda$ to $[-0.20, 0]$. And when $\Lambda=0$, it would reduce to vacuum Kerr black hole case.

Firstly, we paid our attention to $\Lambda=-0.01$ with the observer being located on the equatorial plane. We found that the deformation of the shadows became significant when the spin $a$ is big enough. In particular, as $a$ goes up, the left part of the shadow edge would shrink with some photon ``hairs'' appearing around the left edge. Moreover, we found that when the spin $a$ goes to extremality, the hair structure would be very evident. A similar hair structure was also found in \cite{Wang:2021ara}, where the Kerr Black hole shadows in Melvin magnetic field had been studied. It seems to us that the appearance of the ``hair'' structure in the shadow could be a common feature for the black hole in a magnetic field. The photon hairs may be taken as the signatures of  the magnetic fields outside the black holes.  Remarkably, the structure of the photon hairs may encode more detailed information on the magnetic field.

Then, we moved on to the study of the influences of magnetic fields with various strengths on the shadow while keeping the observer unchanged and fixing the spin $a=0.5$. An interesting feature is that the Kerr black hole shadow becomes more flattened as  the strength of magnetic field increases. The similar behavior has been found for the static black hole case  \cite{Hu:2020usx}. Next, we  discussed the influences of the magnetic fields on the near extremal Kerr black hole with $a=0.999$. In addition to the hair structure that we just mentioned, we noticed the chaotic motions of the photons,  due to the non-integrability of the null geodesics in the effective spacetime. Moreover, we examined the shadows photographed by non-equatorial observers and discussed the features.

Furthermore,  in subsection \ref{qdd}, we made a quantitative analysis of the deformation of the Kerr black hole shadow due to the QED effects, by introducing six characteristic parameters to describe the shadow. Some interesting quantitative results are presented in Table \ref{table} and Fig. \ref{para}. 

\section*{Acknowledgments}
We thank Peng-Cheng Li for his efforts at the initial state of this work. The work is in part supported by NSFC Grant No.  11735001 and No. 11847241.

\begin{appendix} \label{appen}
\section{Explicit expressions of the effective metric functions}\label{appendixA}

\bea
G_{tt}&=&g_{tt}+\frac{a^2 B^2 \lambda  \left[\left(a^2+r^2-2r\right) (\cos 2\theta+3)^2 \left(a^2 \cos2 \theta+a^2-2 r^2\right)^2-8 \left(r^3-a^2 r\right)^2 (\cos4 \theta-1)\right]}{16 \left(a^2 \cos ^2\theta+r^2\right)^5},\nn\\\\
G_{t \phi}&=&g_{t \phi}+\frac{a B^2 \lambda}{16 (a^2 \cos ^2\theta+r^2)^5}\Bigg[16 r (a^2-r^2) (a^2 \cos ^2\theta+r^2)^2  \left(\frac{8 r^3(a^2+r^2)}{(2a^2 \cos^2\theta+2 r^2)^2}+a^2+r^2-2r\right)\sin ^22\theta\nn\\
&&+\Big(a^4 (r-1) \cos4 \theta+3 a^4 (r-1)+4a^2 (a^2 r-a^2+2r^3+r^2) \cos2 \theta+4 a^2 (2 r+1) r^2+8 r^5\Big)\nn\\
&&\cdot \sin ^2\theta(a^2+r^2-2r) (\cos2 \theta+3) (a^2 \cos 2 \theta+a^2-2 r^2)\Bigg],\nn\\\\
G_{rr}&=&g_{rr}+\frac{B^2 \lambda}{16} \Bigg[\frac{a^2 (\cos2 \theta+3)^2 \Big(\frac{4 r (a^2+r^2)}{(2r-r^2-a^2)(a^2 \cos2 \theta+a^2+2 r^2)}-1\Big)(a^2 \cos2 \theta+a^2-2 r^2)^2}{(a^2 \cos ^2\theta +r^2)^4}\nn\\
&&-\bigg(128 a^2 r \sin^2\theta (\cos2 \theta+3) \Big(a^4 (r-1) \cos 4\theta+3 a^4 (r-1)+4 a^2 \big(a^2 (r-1)+r^2 (2 r+1)\big) \cos2 \theta \nn\\
&&+4 a^2 (2 r+1) r^2+8 r^5\Big)(a^2 \cos2 \theta +a^2-2 r^2)-4\Big(a^4 (r-1) \sin5 \theta+a^2(3 a^2 r-3a^2+8r^3+4 r^2) \sin3 \theta\nn\\
&&+2 \big(a^4 (r-1)+2 a^2 (2 r+1) r^2+8 r^5\big) \sin\theta \Big)^2\big(a^2 \cos 2 \theta+a^2+2 (r-2) r\big)\Bigg)\nn\\
&&\cdot  (a^2+r^2-2r)^{-1} (a^2 \cos  2 \theta+a^2+2 r^2)^{-5}\Bigg],\nn\\\\
G_{r\theta}&=&g_{r\theta}+2 B^2 \lambda \sin2\theta (r-1)\nn\\
&&-\frac{2 B^2 \lambda}{(a^2+(r-2) r) (a^2 \cos2 \theta+a^2+2 r^2)^5}\Bigg[16 a^2 r^2 (a-r) (a+r) \sin ^3\theta \cos \theta \nn\\
&&\cdot\Big(a^4 (r-1) \cos4 \theta+3 a^4 (r-1)+4 a^2 \big(a^2r-a^2+2r^3+r^2\big) \cos2 \theta+4 a^2 (2 r+1) r^2+8 r^5\Big)\Bigg]\nn\\
&&+\frac{2 B^2 \lambda \sin2 \theta r}{(a^2+r^2-2r) (a^2 \cos2 \theta+a^2+2 r^2)^5} \Bigg[ 2 (r+1) (a^2+r^2-2r)(a^2 \cos 2 \theta +a^2+2 r^2)^4\nn\\
&&-2 \big(3 a^4+6 a^2 r (2 r-1)+(r-10) r^3\big)(a^2 \cos2 \theta+a^2+2 r^2)^3+64 r^3 (a^2+r^2) (a^4+3 r^4)\nn\\
&&-16 r\big(-a^6 (r-1)+a^4 r^2 (r+1)+a^2 r^4 (5 r+3)+r^6 (3 r+11)\big)(a^2 \cos 2 \theta+a^2+2 r^2)\nn\\
&&+4\big(-a^6+7 a^4 r^2+a^2 r^3 (13 r-12)+r^5 (5 r+4)) (a^2 \cos 2 \theta+a^2+2 r^2\big)^2\Bigg],\nn\\
\eea

\clearpage
\bea
G_{\theta\theta}&=&g_{\theta\theta}+B^2 \lambda  \sin ^22 \theta\Bigg[\frac{a^2 r^2 (a^2-r^2)^2 \Big(-\frac{4 r (a^2+r^2)}{(a^2+(r-2) r)(a^2 \cos2 \theta+a^2+2 r^2)}-1\Big)}{(a^2 \cos ^2\theta+r^2)^4}\nn\\
&&+\frac{\csc ^2\theta\big(a^2 \cos2 \theta+a^2+2 (r-2) r\big)\Big(\frac{8 r^3(a^2+r^2)}{(a^2 \cos2 \theta+a^2+2 r^2)^2}+a^2+(r-2) r\Big)^2}{\big(a^2+(r-2) r\big)(a^2 \cos2 \theta+a^2+2 r^2)}\nn\\
&&+\frac{32\Big(\frac{8 a^2 r^9-8 a^6 r^5}{a^2+(r-2) r}-a^2 r^2 (a-r) (a+r) (a^2 \cos2 \theta+a^2+2 r^2)^2\Big)}{(a^2 \cos2 \theta+a^2+2 r^2)^5}\Bigg],\nn\\\\
G_{\phi \phi}&=&g_{\phi \phi}+\frac{B^2 \lambda}{16 (a^2 \cos ^2\theta+r^2)^5}  \Bigg[16 \sin ^22 \theta \Big(\frac{8 r^3(a^2+r^2)}{(a^2 \cos2 \theta+a^2+2 r^2)^2}+a^2+(r-2) r\Big)^2(a^2 \cos ^2\theta +r^2)^4\nn\\
&&+\big(a^2+(r-2) r\big) \sin ^4\theta\Big(a^4 (r-1) \cos4 \theta+3 a^4 (r-1)+4 a^2\big(a^2 r-a^2+r^2 2 r^3+r^2\big) \cos 2 \theta\nn\\
&&+4 a^2 (2 r+1) r^2+8 r^5\Big)^2\Bigg].
\eea
\end{appendix}


\begin{thebibliography}{10}

\bibitem{LIGOScientific:2016aoc}
{\bfseries LIGO Scientific, Virgo} Collaboration, B.~P. Abbott {\em et~al.},
  ``{Observation of Gravitational Waves from a Binary Black Hole Merger},''
  \href{http://dx.doi.org/10.1103/PhysRevLett.116.061102}{{\em Phys. Rev.
  Lett.} {\bfseries 116} no.~6, (2016) 061102},
  \href{http://arxiv.org/abs/1602.03837}{{\ttfamily arXiv:1602.03837 [gr-qc]}}.

\bibitem{EventHorizonTelescope:2019dse}
{\bfseries Event Horizon Telescope} Collaboration, K.~Akiyama {\em et~al.},
  ``{First M87 Event Horizon Telescope Results. I. The Shadow of the
  Supermassive Black Hole},''
  \href{http://dx.doi.org/10.3847/2041-8213/ab0ec7}{{\em Astrophys. J. Lett.}
  {\bfseries 875} (2019) L1}, \href{http://arxiv.org/abs/1906.11238}{{\ttfamily
  arXiv:1906.11238 [astro-ph.GA]}}.

\bibitem{Synge:1966okc}
J.~L. Synge, ``{The Escape of Photons from Gravitationally Intense Stars},''
  \href{http://dx.doi.org/10.1093/mnras/131.3.463}{{\em Mon. Not. Roy. Astron.
  Soc.} {\bfseries 131} no.~3, (1966) 463--466}.

\bibitem{Bardeen:1973tla}
J.~M. Bardeen, ``{Timelike and null geodesics in the Kerr metric},'' in {\em
  {Les Houches Summer School of Theoretical Physics}: {Black Holes}}.
\newblock 1973.

\bibitem{Cunha:2015yba}
P.~V.~P. Cunha, C.~A.~R. Herdeiro, E.~Radu, and H.~F. Runarsson, ``{Shadows of
  Kerr black holes with scalar hair},''
  \href{http://dx.doi.org/10.1103/PhysRevLett.115.211102}{{\em Phys. Rev.
  Lett.} {\bfseries 115} no.~21, (2015) 211102},
  \href{http://arxiv.org/abs/1509.00021}{{\ttfamily arXiv:1509.00021 [gr-qc]}}.

\bibitem{Wei:2019pjf}
S.-W. Wei, Y.-C. Zou, Y.-X. Liu, and R.~B. Mann, ``{Curvature radius and Kerr
  black hole shadow},''
  \href{http://dx.doi.org/10.1088/1475-7516/2019/08/030}{{\em JCAP} {\bfseries
  08} (2019) 030}, \href{http://arxiv.org/abs/1904.07710}{{\ttfamily
  arXiv:1904.07710 [gr-qc]}}.

\bibitem{Li:2020drn}
P.-C. Li, M.~Guo, and B.~Chen, ``{Shadow of a Spinning Black Hole in an
  Expanding Universe},''
  \href{http://dx.doi.org/10.1103/PhysRevD.101.084041}{{\em Phys. Rev. D}
  {\bfseries 101} no.~8, (2020) 084041},
  \href{http://arxiv.org/abs/2001.04231}{{\ttfamily arXiv:2001.04231 [gr-qc]}}.

\bibitem{Cunha:2016bjh}
P.~V.~P. Cunha, J.~Grover, C.~Herdeiro, E.~Radu, H.~Runarsson, and A.~Wittig,
  ``{Chaotic lensing around boson stars and Kerr black holes with scalar
  hair},'' \href{http://dx.doi.org/10.1103/PhysRevD.94.104023}{{\em Phys. Rev.
  D} {\bfseries 94} no.~10, (2016) 104023},
  \href{http://arxiv.org/abs/1609.01340}{{\ttfamily arXiv:1609.01340 [gr-qc]}}.

\bibitem{Guo:2020zmf}
M.~Guo and P.-C. Li, ``{Innermost stable circular orbit and shadow of the $4D$
  Einstein\textendash{}Gauss\textendash{}Bonnet black hole},''
  \href{http://dx.doi.org/10.1140/epjc/s10052-020-8164-7}{{\em Eur. Phys. J. C}
  {\bfseries 80} no.~6, (2020) 588},
  \href{http://arxiv.org/abs/2003.02523}{{\ttfamily arXiv:2003.02523 [gr-qc]}}.

\bibitem{Perlick:2021aok}
V.~Perlick and O.~Y. Tsupko, ``{Calculating black hole shadows: review of
  analytical studies},'' \href{http://arxiv.org/abs/2105.07101}{{\ttfamily
  arXiv:2105.07101 [gr-qc]}}.

\bibitem{Wang:2016wcj}
M.~Wang, S.~Chen, and J.~Jing, ``{Chaos in the motion of a test scalar particle
  coupling to the Einstein tensor in Schwarzschild\textendash{}Melvin black
  hole spacetime},''
  \href{http://dx.doi.org/10.1140/epjc/s10052-017-4792-y}{{\em Eur. Phys. J. C}
  {\bfseries 77} no.~4, (2017) 208},
  \href{http://arxiv.org/abs/1605.09506}{{\ttfamily arXiv:1605.09506 [gr-qc]}}.

\bibitem{Wang:2018eui}
M.~Wang, S.~Chen, and J.~Jing, ``{Chaotic shadow of a non-Kerr rotating compact
  object with quadrupole mass moment},''
  \href{http://dx.doi.org/10.1103/PhysRevD.98.104040}{{\em Phys. Rev. D}
  {\bfseries 98} no.~10, (2018) 104040},
  \href{http://arxiv.org/abs/1801.02118}{{\ttfamily arXiv:1801.02118 [gr-qc]}}.

\bibitem{Hou:2021okc}
Y.~Hou, M.~Guo, and B.~Chen, ``{Revisiting the shadow of braneworld black
  holes},'' \href{http://dx.doi.org/10.1103/PhysRevD.104.024001}{{\em Phys.
  Rev. D} {\bfseries 104} no.~2, (2021) 024001},
  \href{http://arxiv.org/abs/2103.04369}{{\ttfamily arXiv:2103.04369 [gr-qc]}}.

\bibitem{KIczek:2021vlc}
B.~KIczek and M.~Rogatko, ``{Axion-like dark matter clouds around rotating
  black holes},'' \href{http://arxiv.org/abs/2106.01565}{{\ttfamily
  arXiv:2106.01565 [hep-th]}}.

\bibitem{Cardoso:2019rvt}
V.~Cardoso and P.~Pani, ``{Testing the nature of dark compact objects: a status
  report},'' \href{http://dx.doi.org/10.1007/s41114-019-0020-4}{{\em Living
  Rev. Rel.} {\bfseries 22} no.~1, (2019) 4},
  \href{http://arxiv.org/abs/1904.05363}{{\ttfamily arXiv:1904.05363 [gr-qc]}}.

\bibitem{Contreras:2021yxe}
E.~Contreras, J.~Ovalle, and R.~Casadio, ``{Gravitational decoupling for
  axially symmetric systems and rotating black holes},''
  \href{http://dx.doi.org/10.1103/PhysRevD.103.044020}{{\em Phys. Rev. D}
  {\bfseries 103} no.~4, (2021) 044020},
  \href{http://arxiv.org/abs/2101.08569}{{\ttfamily arXiv:2101.08569 [gr-qc]}}.

\bibitem{Chowdhuri:2020ipb}
A.~Chowdhuri and A.~Bhattacharyya, ``{Shadow analysis for rotating black holes
  in the presence of plasma for an expanding universe},''
  \href{http://arxiv.org/abs/2012.12914}{{\ttfamily arXiv:2012.12914 [gr-qc]}}.

\bibitem{Gralla:2017ufe}
S.~E. Gralla, A.~Lupsasca, and A.~Strominger, ``{Observational Signature of
  High Spin at the Event Horizon Telescope},''
  \href{http://dx.doi.org/10.1093/mnras/sty039}{{\em Mon. Not. Roy. Astron.
  Soc.} {\bfseries 475} no.~3, (2018) 3829--3853},
  \href{http://arxiv.org/abs/1710.11112}{{\ttfamily arXiv:1710.11112
  [astro-ph.HE]}}.

\bibitem{Guo:2018kis}
M.~Guo, N.~A. Obers, and H.~Yan, ``{Observational signatures of near-extremal
  Kerr-like black holes in a modified gravity theory at the Event Horizon
  Telescope},'' \href{http://dx.doi.org/10.1103/PhysRevD.98.084063}{{\em Phys.
  Rev. D} {\bfseries 98} no.~8, (2018) 084063},
  \href{http://arxiv.org/abs/1806.05249}{{\ttfamily arXiv:1806.05249 [gr-qc]}}.

\bibitem{Yan:2019etp}
H.~Yan, ``{Influence of a plasma on the observational signature of a high-spin
  Kerr black hole},'' \href{http://dx.doi.org/10.1103/PhysRevD.99.084050}{{\em
  Phys. Rev. D} {\bfseries 99} no.~8, (2019) 084050},
  \href{http://arxiv.org/abs/1903.04382}{{\ttfamily arXiv:1903.04382 [gr-qc]}}.

\bibitem{Guo:2019lur}
M.~Guo, S.~Song, and H.~Yan, ``{Observational signature of a near-extremal
  Kerr-Sen black hole in the heterotic string theory},''
  \href{http://dx.doi.org/10.1103/PhysRevD.101.024055}{{\em Phys. Rev. D}
  {\bfseries 101} no.~2, (2020) 024055},
  \href{http://arxiv.org/abs/1911.04796}{{\ttfamily arXiv:1911.04796 [gr-qc]}}.

\bibitem{Gralla:2019xty}
S.~E. Gralla, D.~E. Holz, and R.~M. Wald, ``{Black Hole Shadows, Photon Rings,
  and Lensing Rings},''
  \href{http://dx.doi.org/10.1103/PhysRevD.100.024018}{{\em Phys. Rev. D}
  {\bfseries 100} no.~2, (2019) 024018},
  \href{http://arxiv.org/abs/1906.00873}{{\ttfamily arXiv:1906.00873
  [astro-ph.HE]}}.

\bibitem{Peng:2021osd}
J.~Peng, M.~Guo, and X.-H. Feng, ``{Observational Signature and Additional
  Photon Rings of Asymmetric Thin-shell Wormhole},''
  \href{http://arxiv.org/abs/2102.05488}{{\ttfamily arXiv:2102.05488 [gr-qc]}}.

\bibitem{Johannsen:2013vgc}
T.~Johannsen, ``{Photon Rings around Kerr and Kerr-like Black Holes},''
  \href{http://dx.doi.org/10.1088/0004-637X/777/2/170}{{\em Astrophys. J.}
  {\bfseries 777} (2013) 170},
  \href{http://arxiv.org/abs/1501.02814}{{\ttfamily arXiv:1501.02814
  [astro-ph.HE]}}.

\bibitem{Guo:2020qwk}
M.~Guo and S.~Gao, ``{Universal Properties of Light Rings for Stationary
  Axisymmetric Spacetimes},''
  \href{http://dx.doi.org/10.1103/PhysRevD.103.104031}{{\em Phys. Rev. D}
  {\bfseries 103} no.~10, (2021) 104031},
  \href{http://arxiv.org/abs/2011.02211}{{\ttfamily arXiv:2011.02211 [gr-qc]}}.

\bibitem{Zeng:2020dco}
X.-X. Zeng, H.-Q. Zhang, and H.~Zhang, ``{Shadows and photon spheres with
  spherical accretions in the four-dimensional Gauss\textendash{}Bonnet black
  hole},'' \href{http://dx.doi.org/10.1140/epjc/s10052-020-08449-y}{{\em Eur.
  Phys. J. C} {\bfseries 80} no.~9, (2020) 872},
  \href{http://arxiv.org/abs/2004.12074}{{\ttfamily arXiv:2004.12074 [gr-qc]}}.

\bibitem{Zeng:2021dlj}
X.-X. Zeng, G.-P. Li, and K.-J. He, ``{The shadows and observational appearance
  of a noncommutative black hole surrounded by various profiles of
  accretions},'' \href{http://arxiv.org/abs/2106.14478}{{\ttfamily
  arXiv:2106.14478 [hep-th]}}.

\bibitem{Gan:2021pwu}
Q.~Gan, P.~Wang, H.~Wu, and H.~Yang, ``{Photon spheres and spherical accretion
  image of a hairy black hole},''
  \href{http://dx.doi.org/10.1103/PhysRevD.104.024003}{{\em Phys. Rev. D}
  {\bfseries 104} no.~2, (2021) 024003},
  \href{http://arxiv.org/abs/2104.08703}{{\ttfamily arXiv:2104.08703 [gr-qc]}}.

\bibitem{Peng:2020wun}
J.~Peng, M.~Guo, and X.-H. Feng, ``{Influence of quantum correction on black
  hole shadows, photon rings, and lensing rings},''
  \href{http://dx.doi.org/10.1088/1674-1137/ac06bb}{{\em Chin. Phys. C}
  {\bfseries 45} no.~8, (2021) 085103},
  \href{http://arxiv.org/abs/2008.00657}{{\ttfamily arXiv:2008.00657 [gr-qc]}}.

\bibitem{Zhang:2019glo}
M.~Zhang and M.~Guo, ``{Can shadows reflect phase structures of black
  holes?},'' \href{http://dx.doi.org/10.1140/epjc/s10052-020-8389-5}{{\em Eur.
  Phys. J. C} {\bfseries 80} no.~8, (2020) 790},
  \href{http://arxiv.org/abs/1909.07033}{{\ttfamily arXiv:1909.07033 [gr-qc]}}.

\bibitem{Cai:2021fpr}
X.-C. Cai and Y.-G. Miao, ``{Can shadows connect black hole
  microstructures?},'' \href{http://arxiv.org/abs/2101.10780}{{\ttfamily
  arXiv:2101.10780 [gr-qc]}}.

\bibitem{Han:2018ooi}
S.-Z. Han, J.~Jiang, M.~Zhang, and W.-B. Liu, ``{Photon sphere and phase
  transition of d-dimensional (d \ensuremath{\geq} 5) charged
  Gauss\textendash{}Bonnet AdS black holes},''
  \href{http://dx.doi.org/10.1088/1572-9494/aba259}{{\em Commun. Theor. Phys.}
  {\bfseries 72} no.~10, (2020) 105402},
  \href{http://arxiv.org/abs/1812.11862}{{\ttfamily arXiv:1812.11862 [gr-qc]}}.

\bibitem{Wang:2020emr}
X.~Wang, P.-C. Li, C.-Y. Zhang, and M.~Guo, ``{Novel shadows from the
  asymmetric thin-shell wormhole},''
  \href{http://dx.doi.org/10.1016/j.physletb.2020.135930}{{\em Phys. Lett. B}
  {\bfseries 811} (2020) 135930},
  \href{http://arxiv.org/abs/2007.03327}{{\ttfamily arXiv:2007.03327 [gr-qc]}}.

\bibitem{Guo:2019pte}
M.~Guo, P.-C. Li, and B.~Chen, ``{Photon Emission Near Myers-Perry Black Holes
  in the Large Dimension Limit},''
  \href{http://dx.doi.org/10.1103/PhysRevD.101.024054}{{\em Phys. Rev. D}
  {\bfseries 101} no.~2, (2020) 024054},
  \href{http://arxiv.org/abs/1911.08814}{{\ttfamily arXiv:1911.08814 [gr-qc]}}.

\bibitem{Li:2020val}
P.-C. Li, M.~Guo, and B.~Chen, ``{High spin expansion for null geodesics},''
  \href{http://dx.doi.org/10.1088/1361-6382/abd860}{{\em Class. Quant. Grav.}
  {\bfseries 38} no.~6, (2021) 065008},
  \href{http://arxiv.org/abs/2006.05153}{{\ttfamily arXiv:2006.05153 [gr-qc]}}.

\bibitem{Yan:2021yuo}
H.~Yan, M.~Guo, and B.~Chen, ``{Observability of Zero-angular-momentum Sources
  Near Kerr Black Holes},'' \href{http://arxiv.org/abs/2104.07889}{{\ttfamily
  arXiv:2104.07889 [gr-qc]}}.

\bibitem{Cardoso:2008bp}
V.~Cardoso, A.~S. Miranda, E.~Berti, H.~Witek, and V.~T. Zanchin, ``{Geodesic
  stability, Lyapunov exponents and quasinormal modes},''
  \href{http://dx.doi.org/10.1103/PhysRevD.79.064016}{{\em Phys. Rev. D}
  {\bfseries 79} (2009) 064016},
  \href{http://arxiv.org/abs/0812.1806}{{\ttfamily arXiv:0812.1806 [hep-th]}}.

\bibitem{Konoplya:2007yy}
R.~A. Konoplya and R.~D.~B. Fontana, ``{Quasinormal modes of black holes
  immersed in a strong magnetic field},''
  \href{http://dx.doi.org/10.1016/j.physletb.2007.10.065}{{\em Phys. Lett. B}
  {\bfseries 659} (2008) 375--379},
  \href{http://arxiv.org/abs/0707.1156}{{\ttfamily arXiv:0707.1156 [hep-th]}}.

\bibitem{Yang:2021zqy}
H.~Yang, ``{Relating Black Hole Shadow to Quasinormal Modes for Rotating Black
  Holes},'' \href{http://dx.doi.org/10.1103/PhysRevD.103.084010}{{\em Phys.
  Rev. D} {\bfseries 103} no.~8, (2021) 084010},
  \href{http://arxiv.org/abs/2101.11129}{{\ttfamily arXiv:2101.11129 [gr-qc]}}.

\bibitem{Li:2021zct}
P.-C. Li, T.-C. Lee, M.~Guo, and B.~Chen, ``{EQNM/UFPO correspondence for
  Kerr-Newman black hole},'' \href{http://arxiv.org/abs/2105.14268}{{\ttfamily
  arXiv:2105.14268 [gr-qc]}}.

\bibitem{Konoplya:2006qr}
R.~A. Konoplya, ``{Particle motion around magnetized black holes:
  Preston-Poisson space-time},''
  \href{http://dx.doi.org/10.1103/PhysRevD.74.124015}{{\em Phys. Rev. D}
  {\bfseries 74} (2006) 124015},
  \href{http://arxiv.org/abs/gr-qc/0610082}{{\ttfamily arXiv:gr-qc/0610082}}.

\bibitem{Konoplya:2006gg}
R.~A. Konoplya, ``{Magnetized black hole as a gravitational lens},''
  \href{http://dx.doi.org/10.1016/j.physletb.2006.11.018}{{\em Phys. Lett. B}
  {\bfseries 644} (2007) 219--223},
  \href{http://arxiv.org/abs/gr-qc/0608066}{{\ttfamily arXiv:gr-qc/0608066}}.

\bibitem{EventHorizonTelescope:2021bee}
{\bfseries Event Horizon Telescope} Collaboration, K.~Akiyama {\em et~al.},
  ``{First M87 Event Horizon Telescope Results. VII. Polarization of the
  Ring},'' \href{http://dx.doi.org/10.3847/2041-8213/abe71d}{{\em Astrophys. J.
  Lett.} {\bfseries 910} no.~1, (2021) L12},
  \href{http://arxiv.org/abs/2105.01169}{{\ttfamily arXiv:2105.01169
  [astro-ph.HE]}}.

\bibitem{EventHorizonTelescope:2021srq}
{\bfseries Event Horizon Telescope} Collaboration, K.~Akiyama {\em et~al.},
  ``{First M87 Event Horizon Telescope Results. VIII. Magnetic Field Structure
  near The Event Horizon},''
  \href{http://dx.doi.org/10.3847/2041-8213/abe4de}{{\em Astrophys. J. Lett.}
  {\bfseries 910} no.~1, (2021) L13},
  \href{http://arxiv.org/abs/2105.01173}{{\ttfamily arXiv:2105.01173
  [astro-ph.HE]}}.

\bibitem{Lima:2021cgb}
H.~C.~D. Lima, Junior., P.~V.~P. Cunha, C.~A.~R. Herdeiro, and L.~C.~B.
  Crispino, ``{Shadows and lensing of black holes immersed in strong magnetic
  fields},'' \href{http://arxiv.org/abs/2104.09577}{{\ttfamily arXiv:2104.09577
  [gr-qc]}}.

\bibitem{Wang:2021ara}
M.~Wang, S.~Chen, and J.~Jing, ``{Kerr Black hole shadows in Melvin magnetic
  field},'' \href{http://arxiv.org/abs/2104.12304}{{\ttfamily arXiv:2104.12304
  [gr-qc]}}.

\bibitem{Hu:2020usx}
Z.~Hu, Z.~Zhong, P.-C. Li, M.~Guo, and B.~Chen, ``{QED effect on a black hole
  shadow},'' \href{http://dx.doi.org/10.1103/PhysRevD.103.044057}{{\em Phys.
  Rev. D} {\bfseries 103} no.~4, (2021) 044057},
  \href{http://arxiv.org/abs/2012.07022}{{\ttfamily arXiv:2012.07022 [gr-qc]}}.

\bibitem{DeLorenci:2000yh}
V.~A. De~Lorenci, R.~Klippert, M.~Novello, and J.~M. Salim, ``{Light
  propagation in nonlinear electrodynamics},''
  \href{http://dx.doi.org/10.1016/S0370-2693(00)00522-0}{{\em Phys. Lett. B}
  {\bfseries 482} no.~1-3, (2000) 134--140},
  \href{http://arxiv.org/abs/gr-qc/0005049}{{\ttfamily arXiv:gr-qc/0005049}}.

\bibitem{Novello:1999pg}
M.~Novello, V.~A. De~Lorenci, J.~M. Salim, and R.~Klippert, ``{Geometrical
  aspects of light propagation in nonlinear electrodynamics},''
  \href{http://dx.doi.org/10.1103/PhysRevD.61.045001}{{\em Phys. Rev. D}
  {\bfseries 61} (2000) 045001},
  \href{http://arxiv.org/abs/gr-qc/9911085}{{\ttfamily arXiv:gr-qc/9911085}}.
  

\bibitem{Wald:1974np}
R.~M.~Wald,
``Black hole in a uniform magnetic field,''
Phys. Rev. D \textbf{10}, 1680-1685 (1974)
doi:10.1103/PhysRevD.10.1680

\bibitem{Chen:2016tmr}
S.~Chen, M.~Wang, and J.~Jing, ``{Chaotic motion of particles in the
  accelerating and rotating black holes spacetime},''
  \href{http://dx.doi.org/10.1007/JHEP09(2016)082}{{\em JHEP} {\bfseries 09}
  (2016) 082}, \href{http://arxiv.org/abs/1604.02785}{{\ttfamily
  arXiv:1604.02785 [gr-qc]}}.

\end{thebibliography}
\end{document}